\documentclass[amsmath,amssymb,aps,prd,twocolumn,longbibliography,
]{revtex4-1}

\usepackage{graphicx}% Include figure files
\usepackage{dcolumn}% Align table columns on decimal point
\usepackage{bm}% bold math

\begin{document}
\title{
Magnetic hedgehog lattices in noncentrosymmetric metals
}
\author{Shun~Okumura, Satoru~Hayami, Yasuyuki~Kato, and Yukitoshi~Motome}
\affiliation{Department of Applied Physics, the University of Tokyo, Tokyo 113-8656, Japan}

\begin{abstract}
The magnetic hedgehog lattice (HL) is a noncoplanar magnetic texture with a periodic array of magnetic monopoles and anti-monopoles.
Despite phenomenological and numerical studies thus far, there remain open issues on the microscopic origin, especially with respect to the recent experimental findings of two different types of HLs even at zero magnetic field.
Here, we study the stability of the HLs for an effective spin model with long-range interactions arising from itinerant nature of electrons.
By variational calculations and simulated annealing, we find that the HLs are stabilized in the ground state at zero magnetic field by the synergetic effect of the anti-symmetric exchange interactions generated by the spin-orbit coupling and the multiple-spin interactions generated by the spin-charge coupling.
We also clarify the phase diagram in the magnetic fields, which includes topological phase transitions with pair annihilation of the monopoles and anti-monopoles depending on the field directions. 
\end{abstract}

\maketitle
\section{\label{intro}Introduction}
Chirality, often termed as handedness, is a key concept in a broad field of science, ranging from particle physics to biology.
In condensed matter physics, chiral magnetic textures, which break both inversion and mirror symmetries in addition to time-reversal symmetry, have recently attracted considerable attention for potential applications to next-generation electronic devices.
There are a variety of the chiral magnetic textures, such as skyrmion lattices~\cite{Nagaosa2013} and chiral soliton lattices~\cite{Togawa2016}. 
Noncollinear and noncoplanar spin arrangements in these textures generate emergent electromagnetic fields through the Berry phase mechanism, which induce unconventional transport, optical, and magnetoelectric 
properties~\cite{Tokura2010,Mochizuki2015,Tokura2018}.

\begin{figure}[thb]
\centering
\includegraphics[width=\columnwidth,clip]{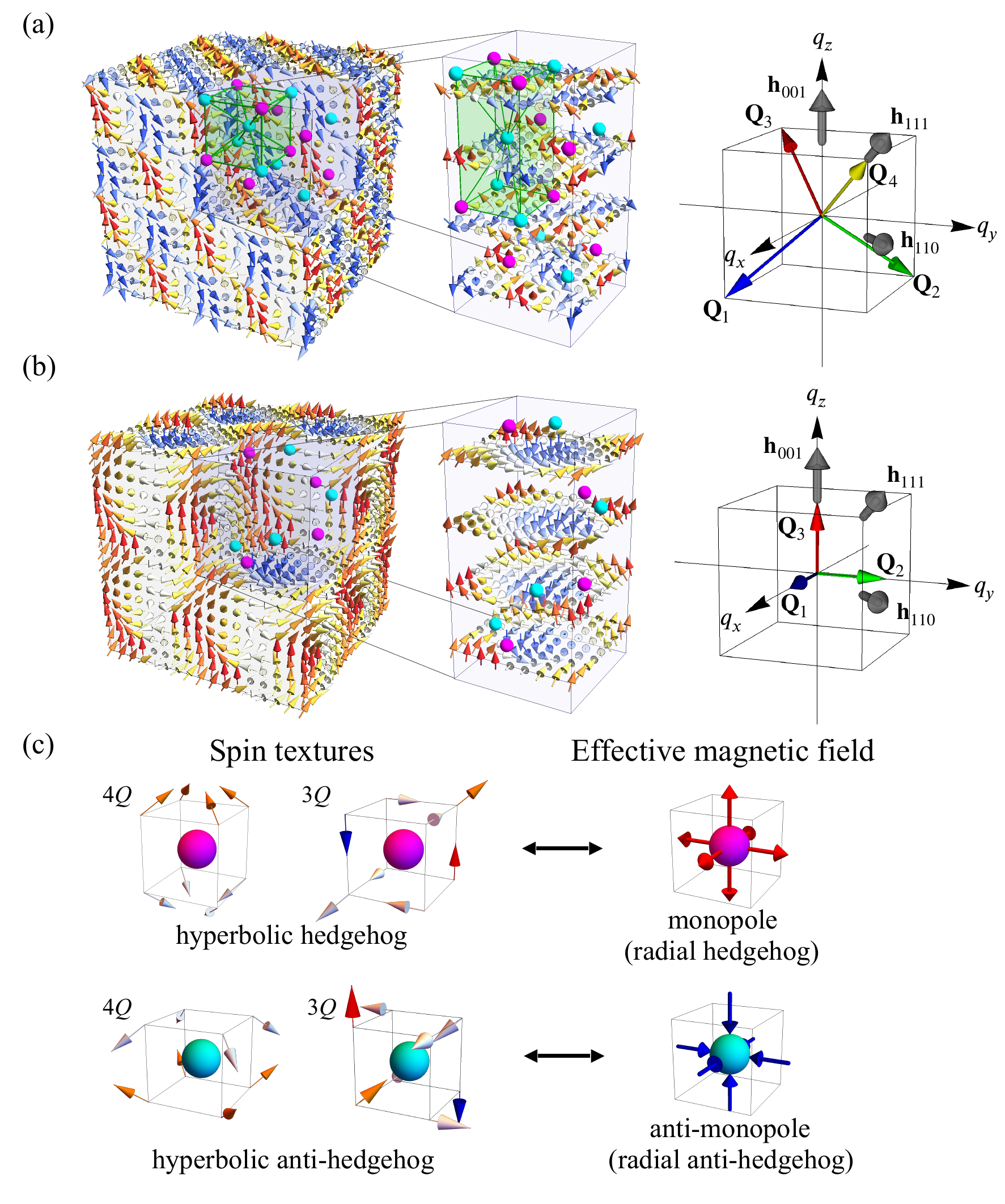}
\caption{
Spin textures of (a) 4$Q$ and (b) 3$Q$ hedgehog lattices obtained by simulated annealing for the model in Eq.~(\ref{eq:Heff}).
The enlarged pictures display the magnetic unit cell with the spin configurations on every two [001] layers for clarity.
The magenta (cyan) balls represent the (anti-)monopoles at the (anti-)hedgehog cores, which locate at the interstitial positions of the cubic lattice sites.
In (a), there are eight monopoles and eight anti-monopoles in the magnetic unit cell, forming two inter-penetrating body-centered-cubic lattices (one of them is shown by the green guides).
Meanwhile, there are four monopoles and four anti-monopoles in (b), which comprise spirals running in the [100], [010], and [001] directions.
The right panels show the ordering vectors for the (a) 4$Q$ and (b) 3$Q$ cases.
The thick arrows (gray) represent the directions of the magnetic field along the [001], [110], and [111] axes.
(c) Correspondences between the spin textures and the effective magnetic fields. 
The cube represents the lattice unit composed of the eight lattice sites surrounding a monopole and an anti-monopole. 
}
\label{f1}
\end{figure}

Recently, a three-dimensional chiral magnetic texture, which is called the hedgehog lattice (HL), was discovered in the $B$20-type compound MnGe~\cite{Tanigaki2015, Kanazawa2017}.  
The magnetic structure is characterized by cubic three wave vectors, and hence, it is referred as the triple-$Q$ hedgehog lattice ($3Q$-HL) [Fig.~\ref{f1}(b)].
The $3Q$-HL has a periodic array of hyperbolic hedgehog and anti-hedgehog spin textures, which generates an emergent magnetic field with a periodic array of radial hedgehogs and anti-hedgehogs regarded as magnetic monopoles and anti-monopoles, as shown in Fig.~\ref{f1}(c)~\cite{Kanazawa2012, Kanazawa2016, Zhang2016}.
The peculiar magnetic field was discussed as a source of the enormous topological Hall effect~\cite{Kanazawa2011} and thermoelectric effect~\cite{Shiomi2013, Fujishiro2018}.
In addition, by a substitution of Ge by Si, the $3Q$-HL changes into a different HL characterized by tetrahedral four wave vectors, dubbed the quadruple-$Q$ hedgehog lattice ($4Q$-HL) [Fig.~\ref{f1}(a)]~\cite{Fujishiro2019}. 
Remarkably, the magnetic periods of these $3Q$- and $4Q$-HLs are very short $\sim2$-$3$~nm, in contrast to most of the skyrmion lattices. 

Such magnetic HLs have been theoretically studied prior to the experimental discovery, e.g., by the Ginzburg-Landau theory~\cite{Binz2006}, variational calculations~\cite{Park2011}, and Monte Carlo (MC) simulations~\cite{Yang2016}.
The variational study for a classical spin model showed that the $3Q$-HL is not stabilized, whereas the $4Q$-HL is obtained in an applied magnetic field~\cite{Park2011}.
The $4Q$-HL in a field was also confirmed by MC simulations~\cite{Yang2016}.
The previous studies, however, do not predict the stable HLs in the absence of magnetic fields, contradicting the experimental observations.
Furthermore, to account for the short-period twist, the localized spin picture requires a large Dzyaloshinskii-Moriya (DM) interaction~\cite{Dzyaloshinskii1958,Moriya1960}, but it was estimated to be very weak~\cite{Gayles2015,Koretsune2015,Kikuchi2016}.
Indeed, recent analyses based on first-principles calculations showed that the stable HLs are not obtained by two spin interactions including the DM interaction~\cite{Grytsiuk2019}. 
The importance of four- and six-spin interactions including spin chirality was also proposed~\cite{Brinker2019,Grytsiuk2020}. 

In this paper, we study the stability of $4Q$- and $3Q$-HLs from a different viewpoint from the previous studies, by taking into account itinerant nature of electrons.
We consider an effective model with long-range exchange interactions originating from the coupling between charge, spin, and orbital degrees of freedom.
By variational calculations and simulated annealing, we show that the model realizes both $4Q$- and $3Q$-HLs at zero field, through the cooperation between the DM-type asymmetric exchange interactions arising from the spin-orbit coupling and the multiple-spin interactions from the spin-charge coupling.
We also study the effect of an applied magnetic field on these HLs. 
Depending on the field directions, we find that the system exhibits multiple phase transitions while changing from the $4Q$- and $3Q$-HLs to the forced ferromagnetic (FFM) state.
Notably, we show that some of them are topological phase transitions with pair annihilation of the monopoles and anti-monopoles.
We demonstrate how the pair annihilation takes place by tracing the positions of the monopoles and anti-monopoles. 

The rest of the paper is organized as follows.
In Sec.~\ref{model}, we introduce the effective spin model derived from an itinerant electron model.
In Sec.~\ref{method}, we describe the methods that we use in this study to investigate the ground state of the effective spin model.
In Sec.~\ref{result:VC}, we show the phase diagram at zero field including the HLs.
In Sec.~\ref{result:SA}, we show the phase diagram in magnetic fields applied in three symmetric directions.
In Sec.~\ref{discuss}, we discuss field-induced topological phase transitions caused by pair annihilation of monopoles and anti-monopoles. 
Section~\ref{summary} is devoted to the summary.

\section{\label{model}Model}
In this section, we present the model which we use in the present study.
Starting from an itinerant electron model with spin-charge and spin orbit couplings in Sec.~\ref{model:KLM}, we discuss the effective model with long-ranged exchange interactions induced by the itinerant nature of electrons in Sec.~\ref{model:ESM}.

\subsection{\label{model:KLM}Itinerant electron model}
In order to investigate the microscopic origin of magnetic HLs, we begin with a minimal model including itinerant electrons, an extended Kondo lattice model that describes the coupling between the itinerant electron spins and localized magnetic moments.
While the Kondo lattice model has been studied for $f$ electron systems, where the $f$ electrons comprises the localized moments~\cite{Stewart1984, Gegenwart2008}, we note that it is also regarded as an effective model for the Hubbard-type models, which have been used widely, e.g., for $d$ electron systems, at the level of the mean-field approximation~\cite{Martin2008}.
In the current study, we include an anti-symmetric spin-orbit coupling arising from spatial inversion symmetry breaking in noncentrosymmetric systems.
The Hamiltonian in the wave-number representation is given by 
\begin{align}
\mathcal{H} = &\sum_{\mathbf{k}\sigma}(\varepsilon_\mathbf{k}-\mu)c^{\dagger}_{\mathbf{k}\sigma}c^{\;}_{\mathbf{k}\sigma}+J_\mathrm{K}\sum_{\mathbf{k}\mathbf{q}\sigma\sigma'}c^{\dagger}_{\mathbf{k}\sigma}{\boldsymbol \sigma}_{\sigma\sigma'}c^{\;}_{\mathbf{k}+\mathbf{q}\sigma'}\cdot{\mathbf S}_{\mathbf{q}}\nonumber\\
&+\sum_{\mathbf{k}\sigma\sigma'}\mathbf{g}_\mathbf{k}\cdot c^{\dagger}_{\mathbf{k}\sigma}{\boldsymbol \sigma}_{\sigma\sigma'}c^{\;}_{\mathbf{k}\sigma'},
\label{eq:KLM}
\end{align}
where $c^{\dagger}_{\mathbf{k}\sigma}$ ($c^{\;}_{\mathbf{k}\sigma}$) is a creation (annihilation) operator of an itinerant electron with wave vector $\mathbf{k}$ and spin $\sigma=\uparrow$ or $\downarrow$. 
The first term describes the kinetic energy of itinerant electrons; $\varepsilon_\mathbf{k}$ is the energy dispersion and $\mu$ is the chemical potential.
The second term is for the Kondo coupling between itinerant electron spins and localized spin moments; ${\boldsymbol \sigma}=(\sigma^x, \sigma^y, \sigma^z)$ is the vector of Pauli matrices, and $\mathbf{S}_\mathbf{q} = \frac{1}{\sqrt{N}} \sum_l \mathbf{S}_{\mathbf{r}_l} e^{-i\mathbf{q}\cdot\mathbf{r}_l}$ is the Fourier transform of a localized moment $\mathbf{S}_{\mathbf{r}_l}=(S^x_{\mathbf{r}_l}, S^y_{\mathbf{r}_l}, S^z_{\mathbf{r}_l})$ defined at site $l$, where $N$ is the number of lattice sites.
For simplicity, $\mathbf{S}_{\mathbf{r}_l}$ is regarded as a classical spin with the length $|\mathbf{S}_{\mathbf{r}_l}|=1$.
$J_\mathrm{K}$ is the exchange coupling constant whose sign is irrelevant for the classical spins.
The last term represents the anti-symmetric spin-orbit coupling induced by spatial inversion symmetry breaking; $\mathbf{g}_\mathbf{k}=(g^x_\mathbf{k}, g^y_\mathbf{k}, g^z_\mathbf{k})$ is called the g-vector, which plays an important role in chiral magnets.
In the following, we consider the model on a simple cubic lattice with the lattice constant being unity for simplicity; noncentrosymmetric nature is effectively taken into account in the g-vector $\mathbf{g}_\mathbf{k}$ with an odd-function of $\mathbf{k}$.

\subsection{\label{model:ESM}Effective spin model}
In general, the coupling between itinerant electrons and localized spins generates effective exchange interactions between the localized spins.
For instance, in the strong-coupling case with $J_\mathrm{K}\gg|\varepsilon_\mathbf{k}-\mu|$, an effective ferromagnetic interaction is generated to maximize the kinetic energy of itinerant electrons by aligning neighboring spins, which is called the double-exchange interaction~\cite{Zener1951,Anderson1955}.
On the other hand, in the weak-coupling case with $J_\mathrm{K}\ll|\varepsilon_\mathbf{k}-\mu|$, the effective magnetic interaction becomes long-ranged and oscillating in space, which is called the Ruderman-Kittel-Kasuya-Yosida (RKKY) interaction~\cite{Ruderman1954,Kasuya1956,Yosida1957}. 
In this study, we consider the weak-coupling case of the model in Eq.~(\ref{eq:KLM}) by an effective spin model derived by perturbation expansion in terms of $J_\mathrm{K}$.
Our model includes a higher-order effect of the spin-charge coupling beyond the RKKY interaction discussed in the previous studies~\cite{Akagi2012,Hayami2014,Hayami2017}, and also a DM-type interaction originating from the spin-orbit coupling in the last term in Eq.~(\ref{eq:KLM})~\cite{Hayami2018}.
The Hamiltonian reads
\begin{align}
\mathcal{H} = \sum_{\eta}\Big[&-J\mathbf{S}_{\mathbf{Q}_\eta}\cdot\mathbf{S}_{-\mathbf{Q}_\eta}+\dfrac{K}{N}({\mathbf S}_{\mathbf{Q}_\eta}\cdot{\mathbf S}_{-\mathbf{Q}_\eta})^2\nonumber\\
&-i{\mathbf D}_\eta\cdot{\mathbf S}_{\mathbf{Q}_\eta}\times{\mathbf S}_{-\mathbf{Q}_\eta}\Big]-\sum_{l}\mathbf{h}\cdot\mathbf{S}_{\mathbf{r}_l}.
\label{eq:Heff}
\end{align}
The first term denotes the RKKY interaction, which is derived by the second-order perturbation with respect to $J_\mathrm{K}$~\cite{Ruderman1954,Kasuya1956,Yosida1957}. 
In general, this tends to stabilize a spiral magnetic texture.
The second term is the biquadratic interaction, which is most relevant among the higher-order perturbations with respect to $J_\mathrm{K}$~\cite{Hayami2017}. 
Hereafter, we consider the positive coupling constant $K>0$, which is known to prefer noncollinear and noncoplanar spin configurations~\cite{Akagi2012,Hayami2014,Hayami2017}. 
The third term represents a DM-type interaction arising from the anti-symmetric spin-orbit coupling, which is derived by the second-order perturbation with respect to $J_\mathrm{K}$~\cite{Hayami2018}. 
This also brings a twist in spin textures, and plays a role in not only choosing the chirality but also giving an anisotropy in spin space.
Note that we ignore other anisotropic exchange interactions originating from the anti-symmetric spin-orbit coupling, for simplicity~\cite{Hayami2018}. 
The last term describes the Zeeman coupling to an external magnetic field $\mathbf{h}$.

In Eq.~(\ref{eq:Heff}), all the exchange interactions are long-ranged in real space and specified by particular wave numbers $\mathbf{Q}_\eta$. 
This inherits the itinerant nature of electrons; specifically, the wave vectors $\mathbf{Q}_\eta$ are set by the multiple maxima in the spin-dependent bare susceptibility of itinerant electrons~\cite{Hayami2014,Hayami2017}.
Corresponding to the $3Q$- and $4Q$-HLs, we assume two sets of $\mathbf{Q}_\eta$: One is a set of the tetrahedral wave vectors as $\mathbf{Q}_1=(Q,-Q,-Q)$, $\mathbf{Q}_2=(-Q,Q,-Q)$, $\mathbf{Q}_3=(-Q,-Q,Q)$, and $\mathbf{Q}_4=(Q,Q,Q)$ [Fig.~\ref{f1}(a)], and the other is a set of the cubic wave vectors as $\mathbf{Q}_1=(Q,0,0)$, $\mathbf{Q}_2=(0,Q,0)$, and $\mathbf{Q}_3=(0,0,Q)$, which are orthogonal to each other [Fig.~\ref{f1}(b)].
In the following calculations, we set $Q=\pi/4$ (period of eight lattice sites); we confirm that the following results remain qualitatively the same for different choices of $Q$.
Although the direction of $\mathbf{D}_\eta$ is independent of that of $\mathbf{Q}_\eta$ in general, we assume $\mathbf{D}_\eta \parallel \mathbf{Q}_\eta$ that stabilizes proper-screw type spin textures~\footnote{The perturbation expansion for Eq.~(\ref{eq:KLM}) leads to $\mathbf{D}_\eta \parallel \mathbf{Q}_\eta$~\cite{Hayami2018}}.
We note that the HLs can be composed of superpositions of the proper screws.
The magnetic field $\mathbf{h}$ is applied along the [001], [110], and [111] directions as shown in the right panels of Figs.~\ref{f1}(a) and \ref{f1}(b).
We set the energy scale as $J=1$.
We consider the system with $N=16^3$ spins under periodic boundary conditions.
We confirmed that the following results remain the same for $N=24^3$ spins (not shown here). 

\section{\label{method}Method}
In this section, we present the methods to study the ground state of the model in Eq.~(\ref{eq:Heff}).
At zero magnetic field, we mainly adopt variational calculations by comparing the energy of several different spin states, as introduced in Sec.~\ref{method:VC}.
In addition, we use simulated annealing, which is introduced in Sec.~\ref{method:SA}, not only to confirm the variational results but also to study the ground state in an applied magnetic field where it is difficult to infer the variational states.

\subsection{\label{method:VC}Variational calculations}
In the variational calculations, we consider the following spin textures as the variational states at zero magnetic field. 
The simplest one is given by 
\begin{align}
	\mathbf{S}_{\mathbf{r}_l} \propto \sum^n_{\eta=1}\hat{\mathbf{a}}_\eta\cos\mathcal{Q}_{\eta l},
\label{eq:NCMQ}
\end{align}
where $\hat{\mathbf{a}}_\eta$ is the unit vector parallel to $\mathbf{Q}_\eta$ and $\mathcal{Q}_{\eta l}=\mathbf{Q}_\eta\cdot\mathbf{r}_l+\varphi_\eta$ ($\varphi_\eta$ represents the phase shift); 
$n=1,2,3$ for the $3Q$ case and $n=1,2,3,4$ for the $4Q$ case.
This is a set of nonchiral states that has no energy gain from the DM-type interaction.
Another variational state is a chiral one described as the equal superpositions of proper screws,
\begin{align}
\mathbf{S}_{\mathbf{r}_l} \propto \sum^n_{\eta=1}(\hat{\mathbf{b}}_\eta\sin\mathcal{Q}_{\eta l}+\hat{\mathbf{c}}_\eta\cos\mathcal{Q}_{\eta l}),
\label{eq:CMQ}
\end{align}
where $\hat{\mathbf{b}}_\eta$ and $\hat{\mathbf{c}}_\eta$ are the unit vectors orthogonal to $\hat{\mathbf{a}}_\eta$ and each other ($\hat{\mathbf{a}}_\eta$, $\hat{\mathbf{b}}_\eta$, and $\hat{\mathbf{c}}_\eta$ form a right-handed system).
Note that the $n=3$ ($n=4$) state for $3Q$ ($4Q$) corresponds to the $3Q$($4Q$)-HL shown in Fig.~\ref{f1}(a)[(b)]. 
In addition, we include another variational state called the double-$Q$ chiral stripe (2$Q$-CS) found in the previous study~\cite{Ozawa2016},
\begin{align}
\mathbf{S}_{\mathbf{r}_l} \propto \sqrt{1-u^2}\hat{\mathbf{b}}_1\sin\mathcal{Q}_{1 l}+\sqrt{1-u^2}\hat{\mathbf{c}}_1\cos\mathcal{Q}_{1 l}+u\hat{\mathbf{a}}_1,
\label{eq:CSVC}
\end{align}
where $u=v\sin\mathcal{Q}_{2 l}$.
In the variational calculations, we compare the energy for all the variational states by varying $\varphi_\eta$ from $0$ to $Q$ and $v$ from $0$ to $1$ to find the lowest-energy candidate for the ground state.

\subsection{\label{method:SA}Simulated annealing} 
In the simulated annealing, we numerically find the candidate for the ground state by mean of MC simulation.
We gradually reduce the temperature of the system from $T=1$ to $T=10^{-5}$ with a condition $T_n=10^{-0.1n}$, where $T_n$ is the temperature in the $n$th step.
During the annealing, we spend a total of $10^5-10^6$ MC sweeps by using the standard Metropolis algorithm.
After annealing at a particular value of the field strength $h=|\mathbf{h}|$, we increase or decrease $h$ successively by $\Delta h= 0.01$.
At every shift by $\Delta h$, we heat the system up to $T=10^{-3}$ and cool down again to $T=10^{-5}$ by annealing. 
Carefully comparing the energy by starting from various values of $h$, we map out the magnetic phase diagram. 

For the state obtained by the simulated annealing, we calculate the magnetization per site along the field direction, 
\begin{align}
m=\frac{1}{N}\sum_{l}\mathbf{S}_{\mathbf{r}_l}\cdot\hat{\mathbf{h}}.
\label{eq:magnetization}
\end{align}
where $\hat{\mathbf{h}}$ is the unit vector in the field direction, and the magnetic susceptibility,
\begin{align}
\chi=\frac{m(h+\Delta h)-m(h)}{\Delta h}.
\label{eq:susceptibility}
\end{align}
To identify the multiple-$Q$ magnetic orders, we also calculate the magnetic moment with wave vector $\mathbf{q}$, 
\begin{align}
m_\mathbf{q}=\sqrt{\frac{S(\mathbf{q})}{N}},
\label{eq:magnetic_moment}
\end{align}
where $S(\mathbf{q})$ is the spin structure factor defined by 
\begin{align}
S(\mathbf{q})=\frac{1}{N}\sum_{l,l'}\mathbf{S}_{\mathbf{r}_l}\cdot\mathbf{S}_{\mathbf{r}_{l'}}e^{i\mathbf{q}\cdot(\mathbf{r}_{l}-\mathbf{r}_{l'})}.
\label{eq:ss_factor}
\end{align}

In addition, following Ref.~\cite{Yang2016}, we define the monopole charge in each unit cube by using the fluxes $\boldsymbol{\Omega}_p$ penetrating six square plaquettes of the cube as~\cite{Okumura2019JPSCP}
\begin{align}
Q_\mathrm{m}(\mathbf{r}_\mathrm{c}) = \frac{1}{4\pi}\sum_{p\in\mathrm{unit\;cube}}\boldsymbol{\Omega}_p\cdot\hat{\mathbf{n}}_p,
\label{eq:monopole_charge}
\end{align}
where $\mathbf{r}_\mathrm{c}$ is the center position of the unit cube and $\hat{\mathbf{n}}_p$ is the normal unit vector of the $p$th plaquette pointing outward of the cube. 
We compute the flux $\boldsymbol{\Omega}_p$ by dividing the $p$th plaquette into two triangles and taking the sum of the solid angles of three spins on the two triangles $i=1$ and $2$. 
Each solid angle is calculated by
\begin{align}
\Omega_i = 2\tan^{-1}\left\{\frac{\mathbf{S}_1\cdot\left(\mathbf{S}_2\times\mathbf{S}_3\right)}{1+\mathbf{S}_1\cdot\mathbf{S}_2+\mathbf{S}_2\cdot\mathbf{S}_3+\mathbf{S}_3\cdot\mathbf{S}_1}\right\},
\label{eq:solid_angle}
\end{align}
where $\mathbf{S}_1$, $\mathbf{S}_2$, and $\mathbf{S}_3$ are the three spins on the $i$th 
triangle in the clockwise order viewed from the center of the cube, and the sign of $\Omega_i$ is taken to be the same as that of $\mathbf{S}_1\cdot\left[\mathbf{S}_2\times\mathbf{S}_3\right]$: $\Omega_i\in[-2\pi,2\pi]$.
The flux $\boldsymbol{\Omega}_p$ is defined as a perpendicular vector to the $p$th plaquette as　\begin{align}
\boldsymbol{\Omega}_p = \sum_{i\in p}\Omega_i\hat{\mathbf{n}}_p.
\label{eq:flux}
\end{align} 
By substituting Eq.~(\ref{eq:flux}) into Eq.~(\ref{eq:monopole_charge}), we obtain the monopole charge $Q_\mathrm{m}(\mathbf{r}_\mathrm{c})$.
This quantity detects the monopoles and anti-monopoles as it takes the value of +1 (-1) when a monopole (anti-monopole) exists in the unit cube.
The monopoles and anti-monopoles are connected by a flow of the flux $\boldsymbol{\Omega}_p$ in Eq.~(\ref{eq:flux}).
We compute the total number of monopoles and anti-monopoles in the magnetic unit cell, $N_\mathrm{m}$, as
\begin{align}
N_\mathrm{m}=\sum_{\mathbf{r}_\mathrm{c}\in V_\mathrm{m}}
|Q_\mathrm{m}(\mathbf{r}_\mathrm{c})|,
\label{eq:monopole_number}
\end{align}
where $V_\mathrm{m}$ is the magnetic unit cell ($8^3$ sites in the following calculations).
We also measure the distances between the monopoles and anti-monopoles by using $\mathbf{r}_\mathrm{c}$ where $\mathbf{Q}_\mathrm{m}(\mathbf{r}_\mathrm{c})=\pm1$.
In particular, we compute the minimum distance between the monopole and anti-monopole by
\begin{align}
d_{\rm m} = \mathrm{min}|\mathbf{r}^\mathrm{m}_\mathrm{c} - \mathbf{r}^\mathrm{a}_\mathrm{c}|,
\label{eq:monopole_distance}
\end{align}
where $\mathbf{r}_\mathrm{c}^\mathrm{m}$ and $\mathbf{r}_\mathrm{c}^\mathrm{a}$ denote $\mathbf{r}_\mathrm{c}$ for the monopoles and anti-monopoles.
This is an important quantity for not only monitoring topological phase transitions by pair annihilation between monopoles and anti-monopoles but also understanding the behavior of the net scalar spin chirality introduced below. 
We note, however, that $\mathbf{r}_\mathrm{c}^\mathrm{m(a)}$ gives an approximate position of the (anti-)monopole core within an accuracy of the lattice constant, and $d_\mathrm{m}$ changes discontinuously by definition.

Finally, we calculate the net scalar spin chirality. 
We define the scalar spin chirality at each lattice site ${\mathbf r}_l$ by the sum of spin triple products on four triangles on the $\alpha\beta$ plane ($\alpha, \beta = x,y,z$) as~\cite{Okumura2019JPSCP} 
\begin{align}
\chi_\mathrm{sc}^\gamma({\mathbf r}_l)=\frac{1}{2}\sum_{\alpha\beta\nu_\alpha\nu_\beta}\epsilon^{\alpha\beta\gamma}\nu_\alpha\nu_\beta\mathbf{S}_{\mathbf{r}_l}\cdot(\mathbf{S}_{{\mathbf{r}_l}+\nu_\alpha\hat{\boldsymbol{\delta}}_\alpha}\times\mathbf{S}_{{\mathbf{r}_l}+\nu_\beta\hat{\boldsymbol{\delta}}_\beta}),
\label{eq:scalar_chi_local}
\end{align}
where $\gamma$ is the perpendicular direction to the $\alpha\beta$ plane, $\epsilon^{\alpha\beta\gamma}$ is the Levi-Civita symbol, $\nu_{\alpha(\beta)}=\pm1$, and $\hat{\boldsymbol{\delta}}_{\alpha(\beta)}$ is the unit translation vector in the $\alpha(\beta)$ direction. 
By taking the sum over all the sites and three planes, we obtain the net spin chirality:
\begin{align}
\chi_\mathrm{sc}=\frac1N \sum_{\gamma l} \chi_\mathrm{sc}^\gamma({\mathbf r}_l). 
\label{eq:scalar_chi}
\end{align}
Since Eqs.~(\ref{eq:solid_angle}) and (\ref{eq:scalar_chi_local}) share the spin triple products, $\chi_\mathrm{sc}$ is related with the (oriented) summation of the flux $\boldsymbol{\Omega}_p$ in Eq.~(\ref{eq:flux}). 
As mentioned above, the flows of the flux connect the monopoles and anti-monopoles, and hence, the lengths of the flux flows, which are approximately given by the distances $|\mathbf{r}_\mathrm{c}^\mathrm{m} - \mathbf{r}_\mathrm{c}^\mathrm{a}|$, affect $\chi_\mathrm{sc}$. 
We will discuss such a relation in Sec.~\ref{discuss}.

\section{\label{result:VC}Phase diagram at zero field} 
\begin{figure}[t]
\centering
\includegraphics[width=\columnwidth,clip]{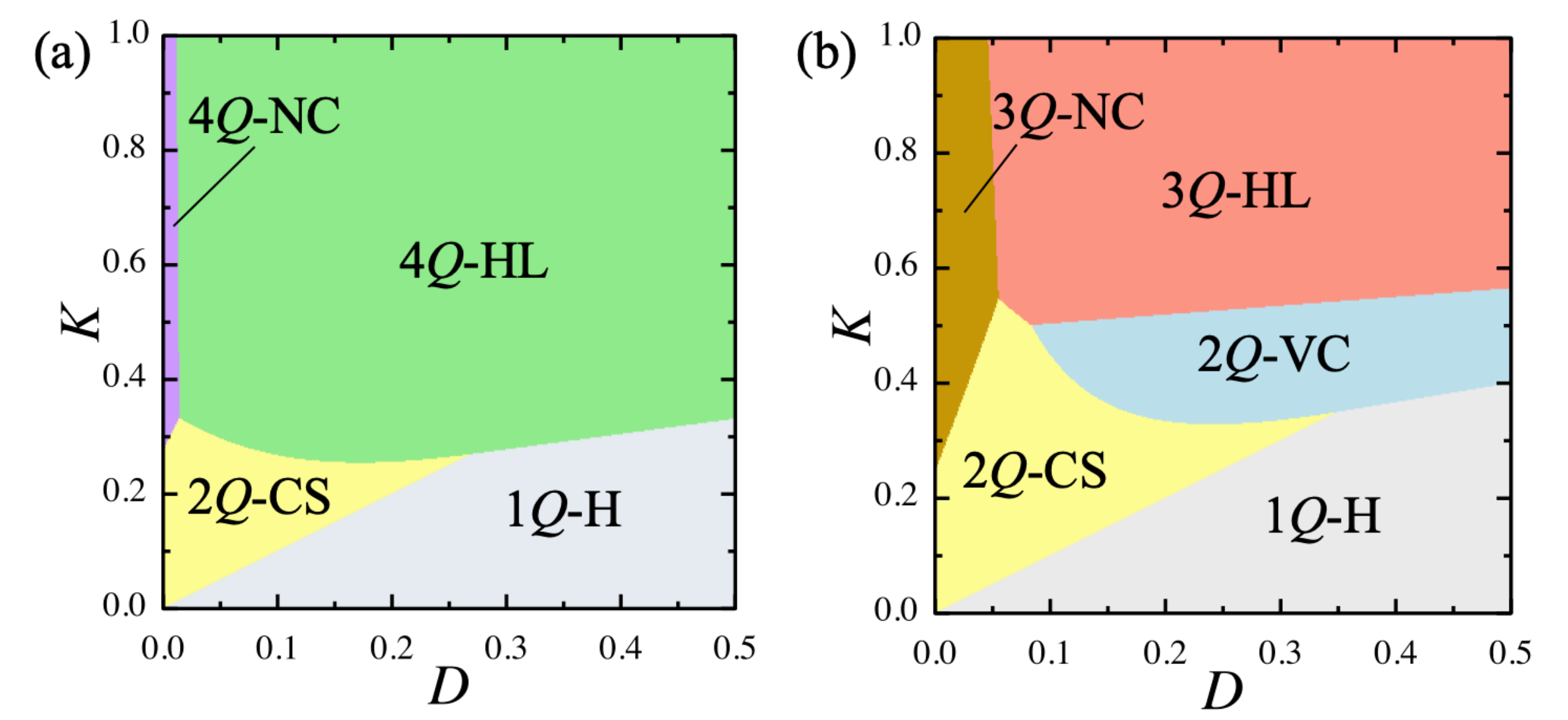}
\caption{
Phase diagrams of the model in Eq.~(\ref{eq:Heff}) at zero field for the (a) 4$Q$ and (b) 3$Q$ cases.
4$Q$($3Q$)-HL, 4$Q$($3Q$)-NC, 2$Q$-VC, 2$Q$-CS, and 1$Q$-H represent the chiral 4$Q$ (3$Q$) hedgehog lattice, the nonchiral 4$Q$ (3$Q$), the chiral 2$Q$ vortex crystal, the 2$Q$ chiral stripe, and the 1$Q$ helical states, respectively.
}
\label{f2}
\end{figure}

First, we show the results in the absence of the magnetic field obtained by the variational calculations in Sec.~\ref{method:VC}.
Figures~\ref{f2}(a) and \ref{f2}(b) display the magnetic phase diagrams for the 4$Q$ and 3$Q$ cases, respectively, while varying $D=|\mathbf{D}_\eta|$ and $K$ in Eq.~(\ref{eq:Heff}).
When $K=0$, a nonzero $D$ stabilizes the chiral $1Q$ helical state ($1Q$-H), which remains stable in the small $K$ region for $D>0$ in both $4Q$ and $3Q$ cases. 
On the other hand, when introducing $K$ with $D=0$, the 2$Q$-CS is stabilized in both cases, but replaced by the nonchiral 4$Q$ and 3$Q$ states in the larger $K$ region. 
Similar sequence of the phase transitions was found in two dimensions~\cite{Ozawa2016,Hayami2017}. 
When $D$ and $K$ are both relevant, however, we find the 4$Q$- and 3$Q$-HLs in the wide parameter range, in addition to a chiral $2Q$ state in the $3Q$ case, which is a Bloch-type vortex crystal ($2Q$-VC)~\cite{Hayami2018}. 
We confirm the stability of these HLs also by the simulated annealing in Sec.~\ref{method:SA}; typical spin configurations for the 4$Q$- and 3$Q$-HLs are presented in Figs.~\ref{f1}(a) and \ref{f1}(b), respectively. 

Thus, our results indicate that the $4Q$- and $3Q$-HLs are stabilized by cooperation between the RKKY interaction, the biquadratic interaction, and the DM-type interaction.
In other words, both spin-charge and spin-orbit couplings play a crucial role in the stabilization of the $4Q$- and $3Q$-HLs. 

From the variational calculations, we find that the stable positions of all the monopoles and anti-monopoles of the $4Q$- and $3Q$-HLs locate not at the lattice sites but at the interstitial positions (centers of unit cubes).
This is concluded for the $3Q$-HL by that the optimized phase shift in Eq.~(\ref{eq:CMQ}) always takes $\varphi_\eta=\pi/8$.
In this case, the eight spins surrounding the (anti-)monopole comprise a hyperbolic (anti-)hedgehog whose north and south poles are in the [111] direction, as shown in Fig.~\ref{f1}(c). 
Meanwhile, for the $4Q$-HL, the set of $\varphi_\eta$ depends on $D$ and $K$ since the four ordering vectors $\bold{Q}_\eta$ are dependent on each other.
In this case, however, the eight spins comprise a hyperbolic (anti-)hedgehog with the north and south poles in the [001] direction. 
In both cases, the (anti-)hedgehogs generates an effective (anti-)monopole field, as shown in Fig.~\ref{f1}(c). 
We deduce that the stable monopoles and anti-monopoles centered at the interstitial positions might be ubiquitous to the systems with fixed spin length on the discrete lattice since their cores are singular points where the spins vanish in the continuum limit. 

\section{\label{result:SA}phase transitions in magnetic fields}
Next, we show the results for the phase diagrams of the model in Eq.~(\ref{eq:Heff}) in the magnetic fields along the [001], [110], and [111] directions obtained by the simulated annealing in Sec.~\ref{method:SA}.
In Secs.~\ref{result:4Q} and \ref{result:3Q}, we present the results for the $4Q$ and $3Q$ cases, respectively.

\subsection{\label{result:4Q}4\boldmath{$Q$} case}

\begin{figure*}[t]
\centering
\includegraphics[width=\linewidth,clip]{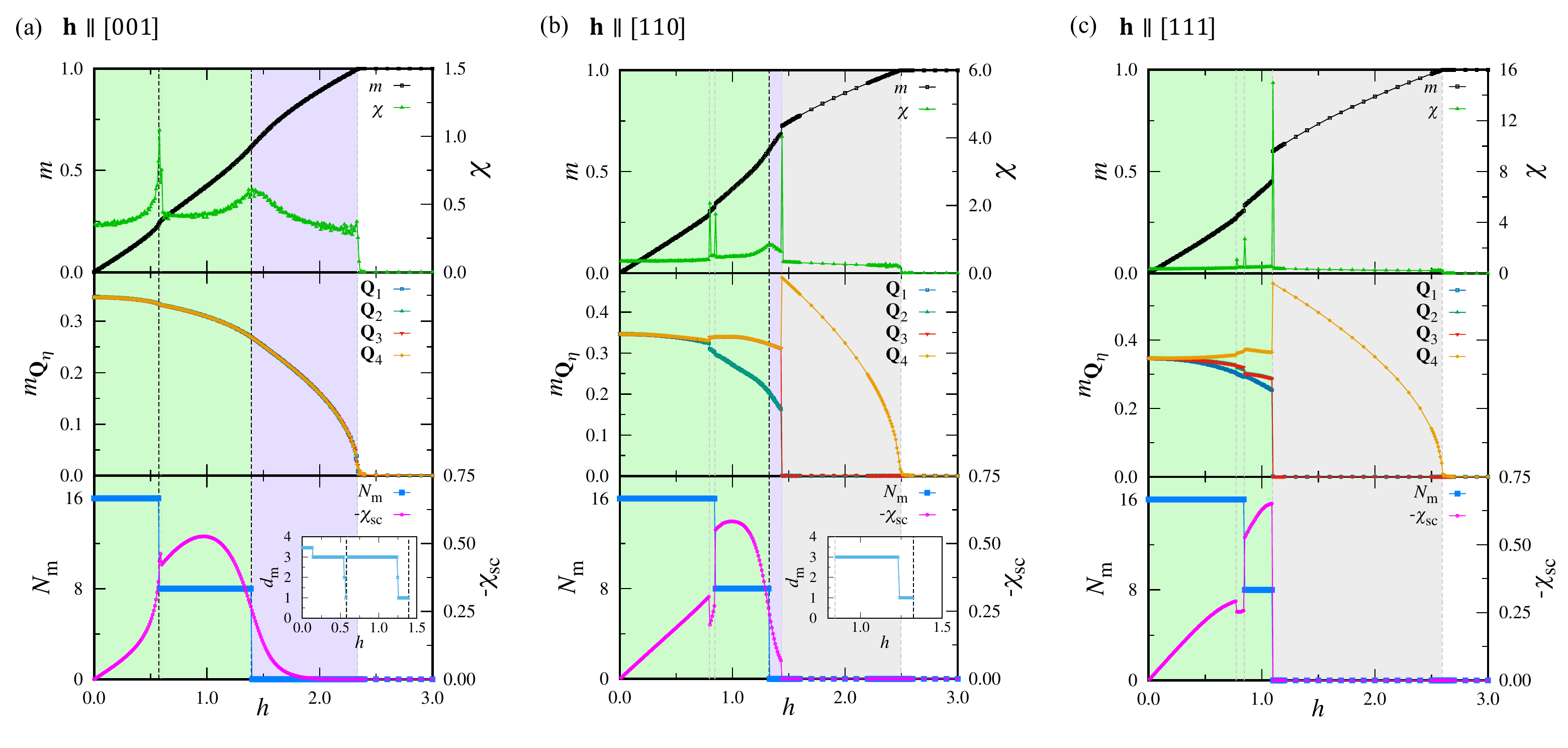}
\caption{
Phase transitions in the magnetic fields along the (a) [001], (b) [110], and (c) [111] directions in the $4Q$ case: the magnetization $m$ in Eq.~(\ref{eq:magnetization}), the magnetic susceptibility $\chi$ in Eq.~(\ref{eq:susceptibility}), the magnetic moments with wave vector $\bold{Q}_\eta$, $m_{\bold{Q}_\eta}$ in Eq.~(\ref{eq:magnetic_moment}), the number of monopoles and anti-monopoles, $N_\mathrm{m}$ in Eq.~(\ref{eq:monopole_number}), and the net scalar spin chirality $\chi_\mathrm{sc}$ in Eq.~(\ref{eq:scalar_chi}) (note that $-\chi_\mathrm{sc}$ is plotted in the figure).
The green, purple, gray, and white regions represent the $4Q$-HLs ($N_\mathrm{m}\neq0$), the noncoplanar $4Q$ states ($N_\mathrm{m}=0$), the $1Q$ conical states, and the FFM state, respectively.
The black-dashed vertical lines represent the topological transitions by pair annihilation of monopoles and anti-monopoles, while the gray ones represent other non-topological phase transitions.
The insets in (a) and (b) show the changes in the minimum distance between monopoles and anti-monopoles, $d_\mathrm{m}$, in Eq.~(\ref{eq:monopole_distance}) when increasing the field before the topological transitions.
See also Figs.~\ref{f5} and \ref{f6}.
}
\label{f3}
\end{figure*}

Let us first discuss the $4Q$ case, whose ordering vectors are shown in Fig.~\ref{f1}(a).
Figure~\ref{f3} summarizes the results for the $4Q$-HL at $D=0.3$ and $K=0.6$.

First, we discuss the results for the [001] field, $\mathbf{h}_{001}=(0,0,h)$, shown in Fig.~\ref{f3}(a).
As plotted in the top panel, the magnetization $m$ shows kinks at $h\simeq0.575$, $1.395$, and $2.335$, and a small jump at $h\simeq 0.595$.
Correspondingly, the magnetic susceptibility $\chi$ shows peaks at $h\simeq0.575$ and $0.595$, a broad hump at $h\simeq1.395$, and a shoulder at $h\simeq2.335$.
These indicate the existence of at least four phase transitions: one at $h\simeq0.595$ is of first order, while the rest three are of second order.
The magnetic moments $m_{\mathbf{Q}_\eta}$ plotted in the middle panel show that the four phases below $h\simeq2.335$ are $4Q$ states with the equal amplitudes for the four $m_{\mathbf{Q}_\eta}$, whereas the phase for $h\gtrsim2.335$ is a FFM state. 
We note that these 4$Q$ states are distinguished by the higher Fourier components of the spin structure factor $S(\mathbf{q})$ (see Appendix~\ref{appendix:A}).

The number of monopoles and anti-monopoles, $N_\mathrm{m}$, is plotted in the bottom panel.
The result shows that $N_{\rm m}$ is halved and vanishes through the second-order phase transitions at $h\simeq0.575$ and $1.395$, respectively (black dashed lines).
As plotted in the inset, the minimum distance between the monopoles and anti-monopoles, $d_\mathrm{m}$, gets shorter from $d_{\rm m}=2\sqrt{3}$ to $1$ and $3$ to $1$ while approaching $h\simeq0.575$ and $1.395$, respectively.
These suggest that the phase transitions are topological ones caused by pair annihilation of monopoles and anti-monopoles. 
We will discuss the details in Sec.~\ref{discuss:4Q}.

In the bottom panel, we also plot the net scalar spin chirality $\chi_{\rm sc}$, which gives rise to the topological Hall effect in itinerant electron systems~\cite{Binz2008}. 
$\chi_{\rm sc}$ rapidly increases before the phase transition at $h\simeq0.575$.
After showing a sharp peak at the phase transition at $h\simeq0.595$, $\chi_{\rm sc}$ exhibits a broad peak at $h \sim 1$, rapidly decreases around the phase transition at $h\simeq 1.395$, and smoothly reduces to zero while approaching the phase transition to a FFM state at $h\simeq 2.335$. 
The change of $\chi_{\rm sc}$ is closely related with the change in the lengths of flows of the flux $\boldsymbol{\Omega}_p$ in Eq.~(\ref{eq:flux}); see Sec.~\ref{discuss:4Q}.

Next, we discuss the results for the [110] field, $\mathbf{h}_{110}=\frac{1}{\sqrt{2}}(h,h,0)$, shown in Fig.~\ref{f3}(b).
As shown in the top panel, $m$ and $\chi$ show jumps and sharp peaks, respectively, at $h\simeq0.795$, $0.845$, and $1.435$.
$m$ also has kinks at $h\simeq1.325$ and $2.495$, where $\chi$ shows a broad hump and a shoulder, respectively.
These indicate the existence of at least five phase transitions: three at $h\simeq0.795$, $0.845$, and $1.435$ are of first order, while the rest two at $h\simeq1.325$ and $2.495$ are of second order. 
$m_{\mathbf{Q}_\eta}$ plotted in the middle panel show that the four phases for $h\lesssim1.435$ are $4Q$ states, the phase for $1.435\lesssim h \lesssim2.495$ is a single-$Q$ ($1Q$) conical state, and that for $h\gtrsim2.495$ is a FFM state.
In the $4Q$ states, the amplitudes of $m_{\mathbf{Q}_\eta}$ are equal at zero field, while they split into two groups for nonzero fields.
We note that the $1Q$ conical phase breaks $C_2$ rotational symmetry spontaneously by choosing one of two equivalent wave vectors $\mathbf{Q}_3$ and $\mathbf{Q}_4$ (we denote the chosen wave vector as $\mathbf{Q}_4$ in the figure).

As shown in the bottom panel of Fig.~\ref{f3}(b), $N_\mathrm{m}$ is halved through the first-order phase transition at $h\simeq0.845$ and vanishes through the second-order one at $h\simeq1.325$. 
As plotted in the inset, $d_{\rm m}$ gets shorter when approaching $h\simeq1.325$, similar to the cases of $\mathbf{h}_{001}$ with $h\simeq 0.575$ and $1.395$.
This also suggests a topological transition by pair annihilation. 
On the other hand, $\chi_{\rm sc}$ has a nonzero value in all the $4Q$ states.
Notably, $\chi_{\rm sc}$ is almost doubled at $h \simeq 0.845$ where $N_{\rm m}$ is halved, and rapidly decreases through the phase transition at $h\simeq 1.325$ where $N_{\rm m}$ vanishes.
We will discuss the relation to the flux flows in Sec.~\ref{discuss:4Q}.

Finally, we discuss the results for the [111] field, $\mathbf{h}_{111}=\frac{1}{\sqrt{3}}(h,h,h)$, shown in Fig.~\ref{f3}(c). 
$m$, $\chi$, and $m_{\mathbf{Q}_\eta}$ in the top and middle panels signal two first-order phase transitions at $h\simeq0.775$ and $0.845$ among the $4Q$-HLs, a first-order one to the $1Q$ conical state at $h\simeq1.095$, and a second-order one to a FFM state at $h\simeq2.595$.
In the $4Q$ states, all four $m_{\mathbf{Q}_\eta}$ become inequivalent for $0.775\lesssim h\lesssim0.845$, while two of them have the same amplitudes for $h\lesssim0.775$ and $0.845\lesssim h\lesssim1.095$. 
This indicates that the $4Q$ state for $0.775 \lesssim h \lesssim 0.845$ has lower symmetry compared to the other two $4Q$ states, while $C_3$ rotational symmetry around the [111] axis ($\parallel \mathbf{Q}_4$) is broken in all three phases except at $h=0$. 

\begin{figure*}[t]
\centering
\includegraphics[width=\linewidth,clip]{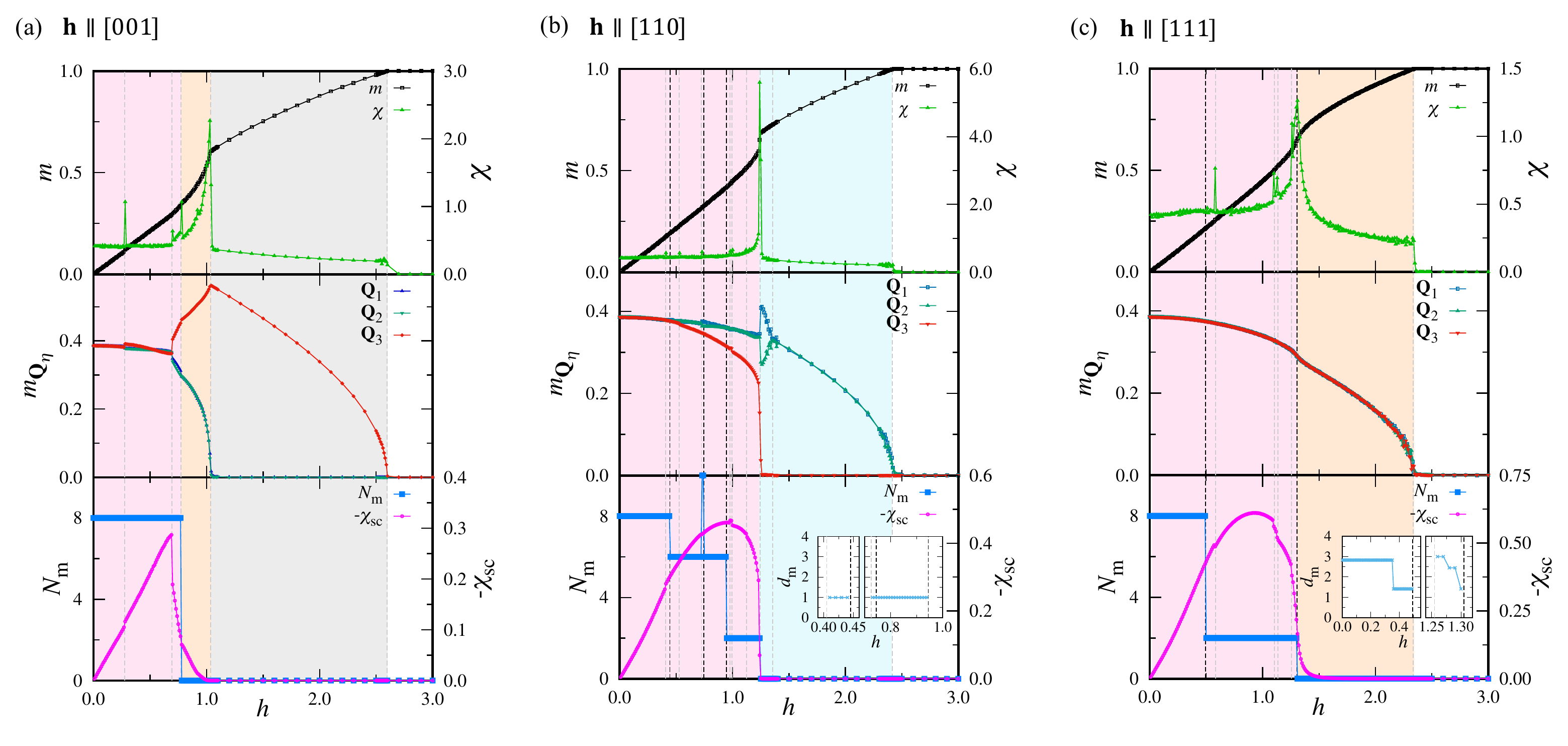}
\caption{
Phase transitions in the magnetic field along the (a) [001], (b) [110], and (c) [111] directions in the $3Q$ case. 
The plotted quantities and the dashed vertical lines for the phase transitions are common to those in Fig.~\ref{f3}. 
The red, orange, blue, gray, and white regions represent the $3Q$-HLs ($N_\mathrm{m}\neq0$), the noncoplanar $3Q$ states ($N_\mathrm{m}=0$), the $2Q$ vortex crystal states, the $1Q$ conical state, and the FFM state, respectively.
See also Figs.~\ref{f7} and \ref{f8} for the changes of $d_{\rm m}$ shown in the insets of the bottom panels of (b) and (c).
}
\label{f4}
\end{figure*}

As plotted in the bottom panel in Fig.~\ref{f3}(c), $N_{\rm m}$ is reduced to half through the first-order phase transition at $h\simeq0.845$. 
This leads to the enhancement of $\chi_{\rm sc}$, similar to the case with $\mathbf{h}_{110}$.
In the [111] field, however, the system does not exhibit a continuous phase transition that might be ascribed to the topological phase transition.
This is presumably due to the fact that the $1Q$ conical state is more stable down to a lower field, compared to the [001] and [110] cases, since the field is applied in parallel to one of the wave vectors, $\mathbf{Q}_4$. 

\subsection{\label{result:3Q}3\boldmath{$Q$} case}

Next, we discuss the $3Q$ case with the ordering vectors shown in Fig.~\ref{f1}(b).
Figure~\ref{f4} summarizes the results for the $3Q$-HL at $D=0.3$ and $K=0.7$. 

First, we discuss the results for the [001] field, $\mathbf{h}_{001}=(0,0,h)$, shown in Fig.~\ref{f4}(a).
As shown in the top and middle panels, $m$, $\chi$, and $m_{\mathbf{Q}_\eta}$ signal at least five phase transitions: first-order ones at $h\simeq0.275$, $0.695$, and $0.775$, and second-order ones at $h\simeq1.035$ and $2.595$.
The four low-field phases for $h\lesssim1.035$ are $3Q$ states with nonzero three $m_{\mathbf{Q}_\eta}$, the phase for $1.035\lesssim h\lesssim 2.595$ is a $1Q$ conical state with only $m_{\mathbf{Q}_3}\neq 0$, and that for $h\gtrsim 2.595$ is a FFM state. 
Furthermore, when we look closer $m_{\mathbf{Q}_\eta}$, we find that $m_{\mathbf{Q}_1}$ becomes inequivalent to $m_{\mathbf{Q}_2}$ at $0.695\lesssim h\lesssim 0.775$, whereas $m_{\mathbf{Q}_1} = m_{\mathbf{Q}_2}$ in the other three $3Q$ states. 
These 3$Q$ states are also distinguished by the higher Fourier components of the spin structure factor $S(\mathbf{q})$ and the structure factor of the local scalar spin chirality (see Appendix~\ref{appendix:B}).

As shown in the bottom panel of Fig.~\ref{f4}(a), $N_{\rm m}$ is unchanged in the three low-field $3Q$ phases, but it vanishes through the first-order phase transition at $h\simeq0.775$.
On the other hand, $\chi_{\rm sc}$ increases in the two low-field phases, while it rapidly decreases in the third phase and vanishes through the second-order phase transition to the $1Q$ conical state at $h\simeq1.035$. 
The change of $\chi_{\rm sc}$ in the $3Q$-HL phases is accounted for by the change in the lengths of the flux flows connecting the monopoles and anti-monopoles, similar to the $4Q$ case in Sec.~\ref{result:4Q} (see Sec.~\ref{discuss:3Q}).

Next, we discuss the results for the [110] field, $\mathbf{h}_{110}=\frac{1}{\sqrt{2}}(h,h,0)$, shown in Fig.~\ref{f4}(b).
As plotted in the top panel, the data of $m$ and $\chi$ signal seven first-order phase transitions at $h\simeq0.405$, $0.525$, $0.725$, $0.975$, $0.995$, $1.125$, and $1.245$, and a second-order phase transition at $h\simeq2.415$. 
In addition, $m_{\mathbf{Q}_\eta}$ in the middle panel and $N_{\rm m}$ in the bottom panel indicate additional phase transitions at $h\simeq0.445$, $0.745$, $0.945$, and $1.355$.
$m_{\mathbf{Q}_\eta}$ shows that all the phases for $h\lesssim 1.245$ are $3Q$ states, the two phases for $1.245\lesssim h\lesssim 2.415$ are $2Q$ states, and the phase for $h\simeq2.415$ is a FFM state. 
We note that $m_{\mathbf{Q}_1}$ becomes inequivalent to $m_{\mathbf{Q}_2}$ in the $3Q$ states for $0.405\lesssim h\lesssim 0.525$ and $0.725\lesssim h\lesssim 0.975$, and the $2Q$ state for $1.245\lesssim h\lesssim 1.355$. 
This indicates spontaneous symmetry breaking by choosing one of the two equivalent wave vectors in these states.

Within the $3Q$ phases for $h\lesssim 1.245$, $N_{\rm m}$ changes in a complicated manner, as plotted in the bottom panel of Fig.~\ref{f4}(b): 
In contrast to the other cases, $N_{\rm m}$ is not reduced monotonically but changes from 8, 6, 10, 6, to 2 stepwisely.
By tracing $d_{\rm m}$ plotted in the insets, we find that the three phase transitions at $h\simeq0.445$, $0.745$, and $0.945$ appear to be topological ones caused by pair annihilation of monopoles and anti-monopoles ($d_{\rm m}$ does not change from $1$ before the transitions since the lattice spacing is larger than the positional changes of monopoles and anti-monopoles; see Sec.~\ref{discuss:3Q} for the details).
The net scalar spin chirality $\chi_{\rm sc}$ is nonzero in all the $3Q$-HLs for $0<h\lesssim 1.245$.
It exhibits a broad peak at $h \sim 1$ and vanishes through the first-order phase transition to the $2Q$ state at $h\simeq1.245$.

Finally, we discuss the results for the [111] field, $\mathbf{h}_{111}=\frac{1}{\sqrt{3}}(h,h,h)$, shown in Fig.~\ref{f4}(c).
In this case, $m$ and $\chi$ plotted in the top panel signal four first-order phase transitions at $h\simeq0.585$, $1.105$, $1.135$, and $1.255$, and two second-order ones at $h\simeq1.305$ and $2.335$. 
In addition, $N_{\rm m}$ in the bottom panel indicates an additional phase transition at $h\simeq0.495$.  
$m_{\mathbf{Q}_\eta}$ in the middle panel shows that all the phases for $h\lesssim2.335$ are $3Q$ states, while the phase for $h\gtrsim2.335$ is a FFM state.
All the $3Q$ phases have the equal amplitudes for the three $m_{\mathbf{Q}_\eta}$; namely, they retain $C_3$ rotational symmetry with respect to the [111] axis.

\begin{figure}[b]
\centering
\includegraphics[width=\columnwidth,clip]{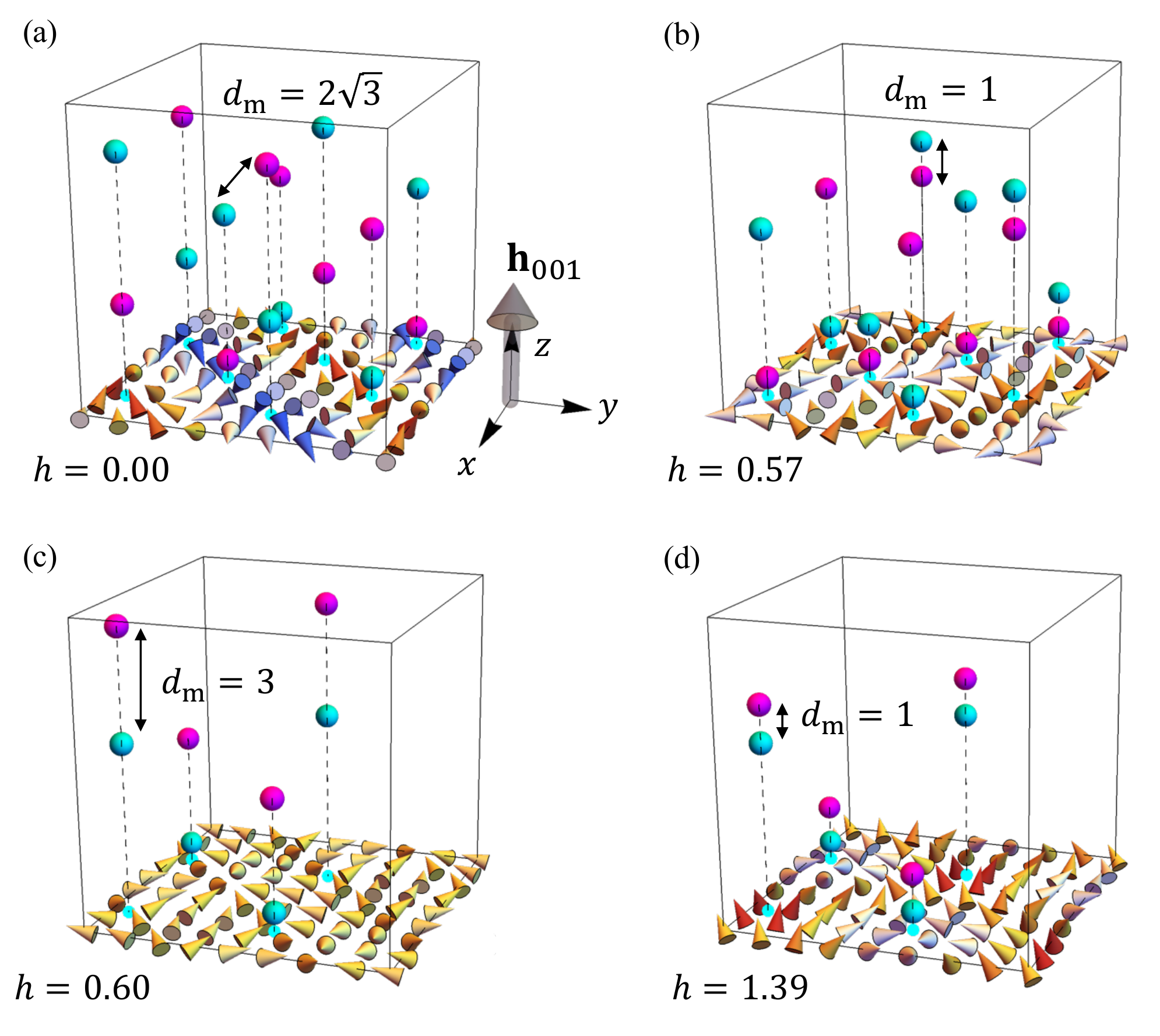}
\caption{
Positions of monopoles (magenta) and anti-monopoles (cyan) in the magnetic unit cell (cube) when approaching the topological transition at $h \simeq 0.575$ and $1.395$ for the [001] field (denoted by the gray arrow) 
in the $4Q$ case: (a) $h=0.00$, (b) $0.57$, (c) $0.60$, and (d) $1.39$.
The arrows at the bottom show the slice of the spin texture on the plane just below some of the monopoles and anti-monopoles.
The black arrows represent the minimum distances between the monopoles and anti-monopoles, $d_{\rm m}$.
The vertical dashed lines and the dots at the bottom end represent the projections onto the bottom plane as guide for the eye.
}
\label{f5}
\end{figure}

As plotted in the bottom panel of Fig.~\ref{f4}(c), $N_{\rm m}$ is nonzero in the $3Q$ phases below $h\simeq 1.305$.
By monitoring $d_{\rm m}$ plotted in the insets, we find that the transitions at $h\simeq0.495$ and $1.305$ appears to be topological ones by pair annihilation of monopoles and anti-monopoles. 
$\chi_\mathrm{sc}$ is nonzero for all the $3Q$-HLs but decreases rapidly through the second-order phase transition at $h\simeq1.305$ where $N_\mathrm{m}$ vanishes.
We will discuss the details in Sec.~\ref{discuss:3Q}.

\section{\label{discuss}topological phase transitions by pair annihilation of monopoles and anti-monopoles}
In Sec.~\ref{result:SA}, we found several phase transitions in the $4Q$- and $3Q$-HL phases where no discontinuous changes are observed in $m$ and $m_{\mathbf{Q}_\eta}$ but $N_{\rm m}$ changes. 
These suggest continuous phase transitions with a topological change caused by pair annihilation of monopoles and anti-monopoles. 
Such topological transitions under the [001] field were discussed for an ansatz of the $3Q$-HL state in the continuum limit~\cite{Zhang2016} and also for a metastable $3Q$-HL in the model in Eq.~(\ref{eq:Heff})~\cite{Okumura2019JPSCP}. 
Our results in Sec.~\ref{result:SA}, however, appear to offer several examples in the ground state for both $4Q$- and $3Q$-HLs.
In this section, we analyze these phase transitions by tracing the positions of monopoles and anti-monopoles in real space.
In Secs.~\ref{discuss:4Q} and \ref{discuss:3Q}, we present the results for the $4Q$ and $3Q$ cases, respectively.

\subsection{\label{discuss:4Q}4\boldmath{$Q$} case}

\begin{figure}[b]
\centering
\includegraphics[width=\columnwidth,clip]{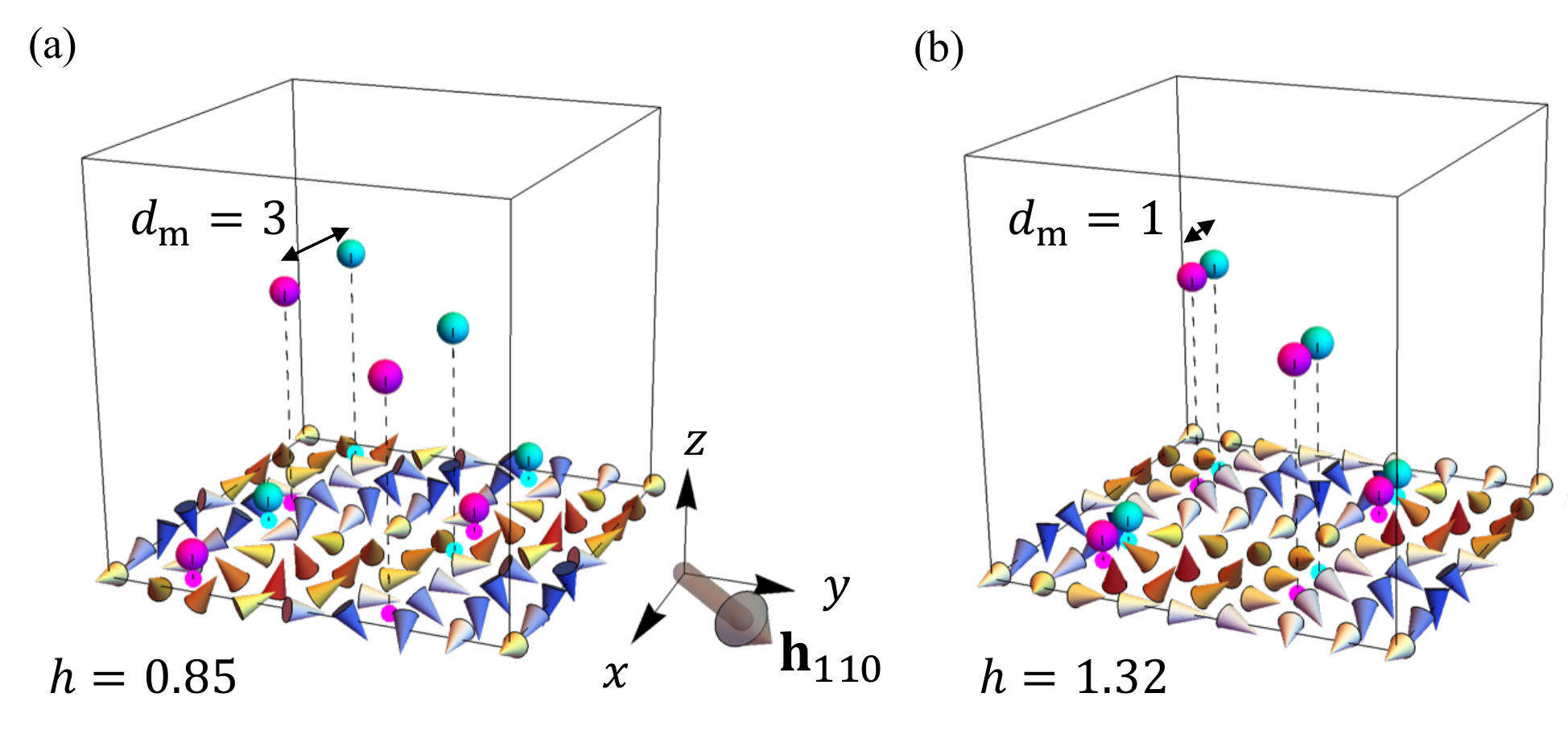}
\caption{
Positions of monopoles and anti-monopoles when approaching the topological transition at $h \simeq 1.325$ for the [110] field in the $4Q$ case: (a) $h=0.85$ and (b) $h=1.30$.
The notations are common to those in Fig.~\ref{f5}.
}
\label{f6}
\end{figure}

In Sec.~\ref{result:4Q}, we found three possible topological phase transitions in the $4Q$-HLs:
Two are at $h\simeq0.575$ and $1.395$ for the [001] field and the other is at $h\simeq1.325$ for the [110] field.
We discuss how the monopoles and anti-monopoles move and pair annihilate as a function of the field strength through each transition.

In the case of the [001] field, $N_{\rm m}$ changes from $16$ to $8$ at $h\simeq0.575$ and from $8$ to $0$ at $h\simeq1.395$, both suggesting four pairs of monopoles and anti-monopoles annihilate simultaneously at the phase transition. 
They are visualized in real space in Fig.~\ref{f5}. 
At zero field, the monopoles and anti-monopoles form two inter-penetrating body-centered-cubic lattices with $d_\mathrm{m}=2\sqrt{3}$ as shown in Fig.~\ref{f5}(a).
While increasing $h$, half of the monopoles and anti-monopoles move toward each other in the field direction, forming four pairs.
When approaching to the critical field, $d_{\rm m}$ given by the four pairs is reduced to $1$ as shown in Fig.~\ref{f5}(b) at $h=0.57$, and then, becomes $0$, which is the pair annihilation at the critical field $h\simeq0.575$.
In the higher-field region, the remaining monopoles and anti-monopoles are paired along the field direction again, as exemplified in Fig.~\ref{f5}(c) at $h=0.60$. 
In this case, $d_{\rm m}$ is reduced from $3$ to $1$ as shown in Fig.~\ref{f5}(d) at $h=1.39$, and finally, becomes $0$ at the critical field $h\simeq1.395$ by the pair annihilation.
The changes of $d_{\rm m}$ were plotted in the inset of the bottom panel in Fig.~\ref{f3}(a). 

\begin{figure}[b]
\centering
\includegraphics[width=\linewidth,clip]{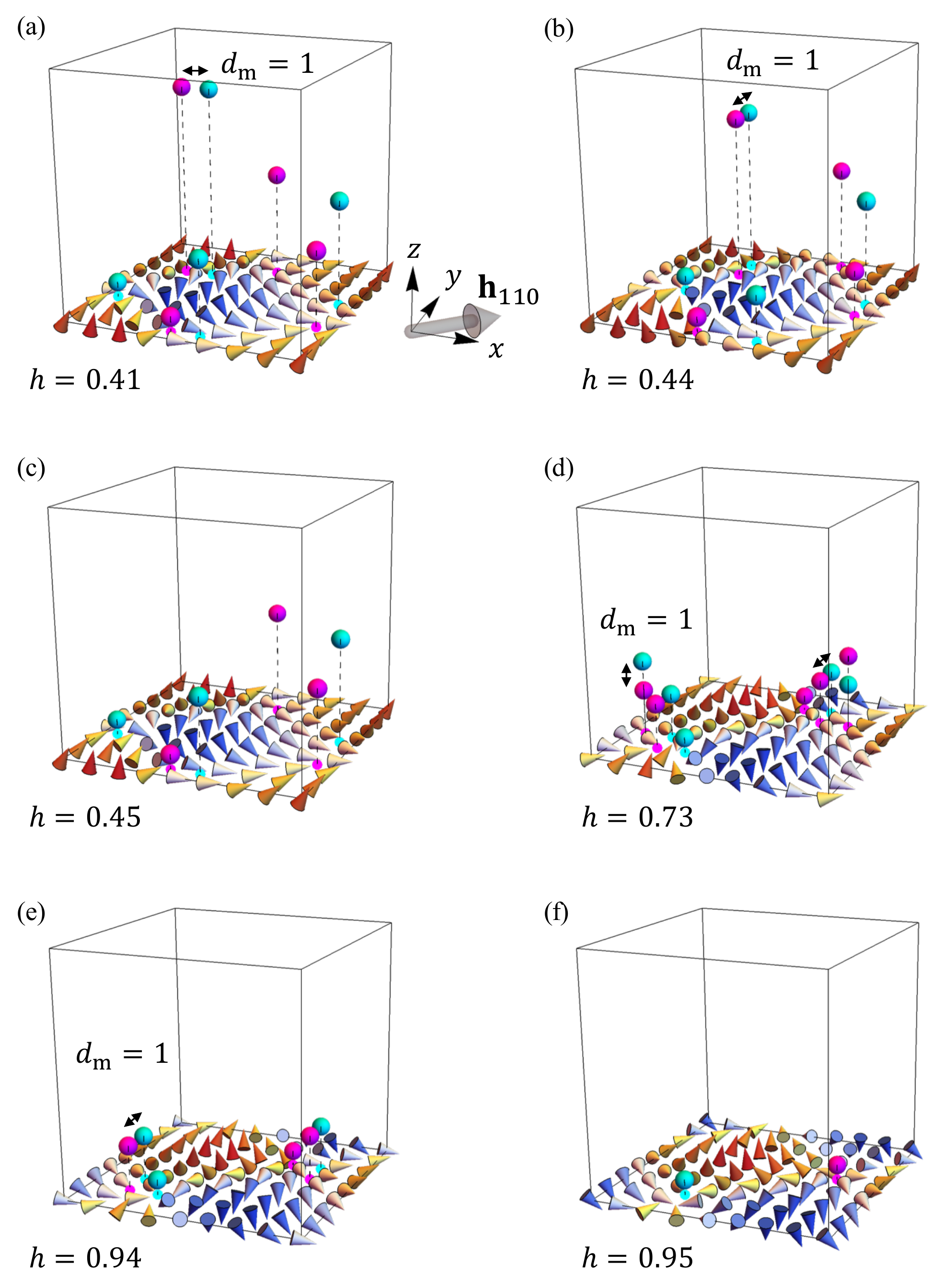}
\caption{
Positions of monopoles and anti-monopoles when approaching the topological transition at $h \simeq 0.445$, 0.745, and 0.945 for the [110] field in the $3Q$ case: (a) $h=0.41$, (b) $h=0.44$, (c) $h=0.45$, (d) $h=0.73$, (e) $h=0.94$, and (f) $h=0.95$.
The notations are common to those in Fig.~\ref{f5}.
}
\label{f7}
\end{figure}

The movement of the monopoles and anti-monopoles explains the behavior of $\chi_\mathrm{sc}$ plotted in the bottom panel of Fig.~\ref{f3}(a). 
When approaching the topological transition at $h\simeq 0.575$ by increasing $h$, $\chi_\mathrm{sc}$ decreases [$-\chi_\mathrm{sc}$ increases in Fig.~\ref{f3}(a)]. 
This is understood by the decrease of $d_\mathrm{m}$ with the flux flows in the same direction of the magnetic field: 
The decrease of $d_\mathrm{m}$ reduces the positive contribution to $\chi_\mathrm{sc}$, which leads to the net decrease in $\chi_\mathrm{sc}$. 
On the other hand, $\chi_\mathrm{sc}$ increases ($-\chi_\mathrm{sc}$ decreases) near the other topological transition at $h\simeq 1.395$. 
This is due to the decrease of $d_\mathrm{m}$ with the flux flows in the opposite direction to the magnetic field.

Similarly, $N_{\rm m}$ changes from $8$ to $0$ for the [110] field through the phase transition at $h\simeq 1.325$.
The change of the positions of monopoles and anti-monopoles is shown in Fig.~\ref{f6}, where $d_{\rm m}$ changes in a similar manner to the case of the [001] field at $h\simeq 1.39$ in Figs.~\ref{f5}(c) and \ref{f5}(d); see also the inset of the bottom panel of Fig.~\ref{f3}(b). 
The only difference from the [001] case is in the direction of collisions.
The corresponding reduction of the lengths of the flux flows is also related to the suppression of $\chi_\mathrm{sc}$ in Fig.~\ref{f3}(b) since the fluxes $\boldsymbol{\Omega}_p$ have the positive component in the opposite direction to the field.

\subsection{\label{discuss:3Q}3\boldmath{$Q$} case}

\begin{figure}[b]
\centering
\includegraphics[width=\linewidth,clip]{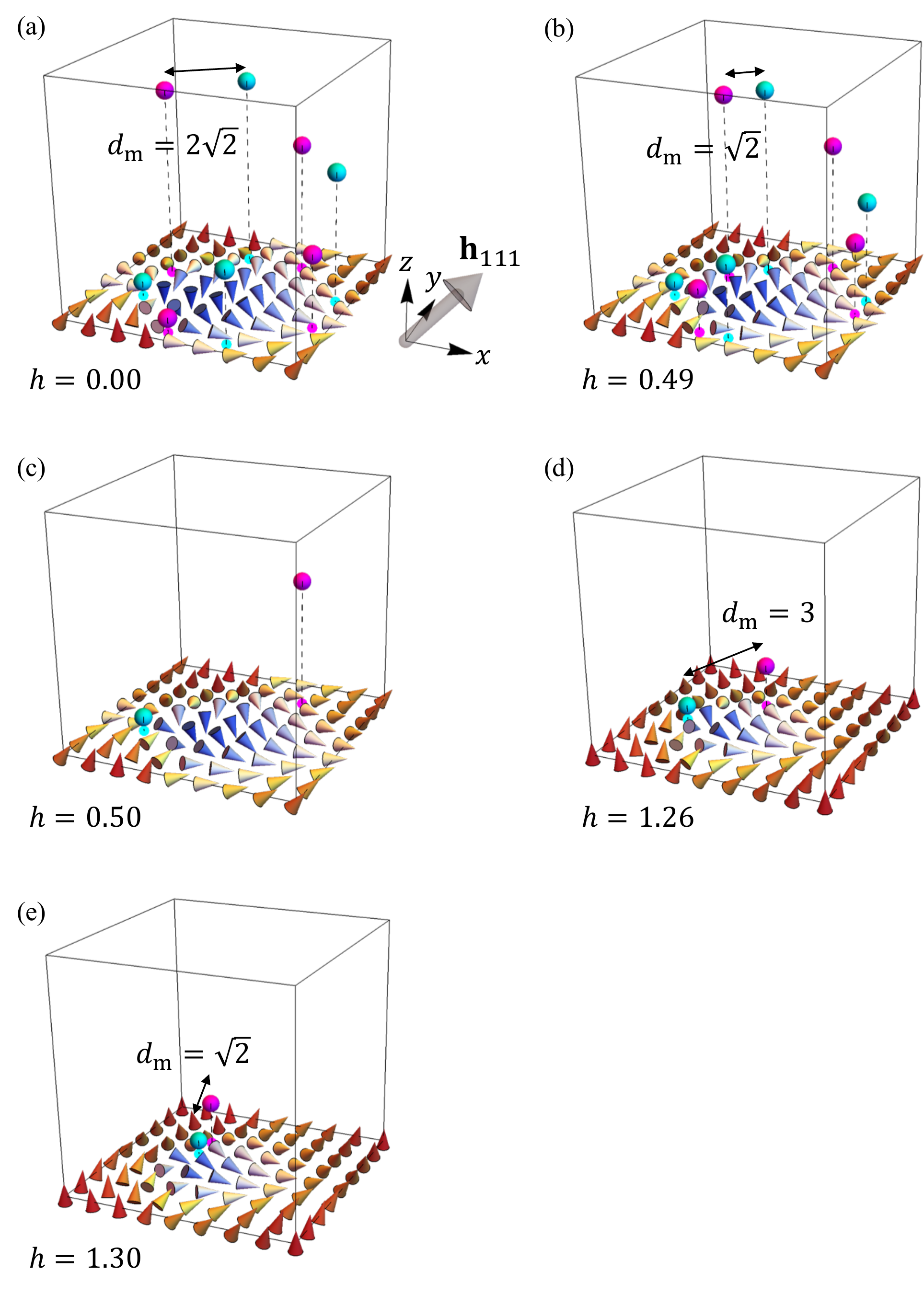}
\caption{
Positions of monopoles and anti-monopoles when approaching the topological transition at $h \simeq 0.495$ and 1.305 for the [111] field in the $3Q$ case: (a) $h=0.00$, (b) $h=0.49$, (c) $h=0.50$, (d) $h=1.26$, and (e) $h=1.30$.
The notations are common to those in Fig.~\ref{f5}.
}
\label{f8}
\end{figure}

In the case of the $3Q$-HLs, we identified totally five possible topological transitions in Sec.~\ref{result:3Q}.
Three of them are at $h\simeq0.445$, $0.745$, and $0.945$ for the [110] field, and the rest two are at $h\simeq0.495$ and $1.305$ for the [111] field. 
Figure~\ref{f7} shows the real-space pictures for the [110] field. 
In a low field, there are totally eight monopoles and anti-monopoles as shown in Fig.~\ref{f7}(a) for $h=0.41$, but one pair annihilates through the transition at $h\simeq0.445$ as shown in Figs.~\ref{f7}(b) and \ref{f7}(c). 
We note that the directions of pairs change within the same $3Q$ state with spontaneous symmetry breaking with respect to the [100] and [010] directions.
In the next topological transition at $h\simeq0.745$, $N_\mathrm{m}$ changes from $10$ to $6$, where two pairs of monopoles and anti-monopoles annihilate as shown in Figs.~\ref{f7}(d) and \ref{f7}(e). 
Through these transitions, $d_{\rm m}$ does not change from $1$, since the distance for the pairs that survive is already $1$ before the transition.
Finally, other two pairs annihilate and $N_\mathrm{m}$ is reduced to 2 at $h\simeq0.945$ as shown in Fig.~\ref{f7}(f).
See also the insets in the bottom panel of Fig.~\ref{f4}(b).

Finally, we present the results for the [111] field in Fig.~\ref{f8}.
In this case, there are four pairs of monopoles and anti-monopoles in the low-field phase, and their distance gets shorter as demonstrated in Figs.~\ref{f8}(a) and \ref{f8}(b).
Through the topological transition at $h\simeq 0.495$, three of four annihilate as shown in Fig.~\ref{f8}(c). 
Finally, the remaining pair gets closer and pair annihilate through the transition at $h\simeq1.305$ as shown in Figs.~\ref{f8}(d) and \ref{f8}(e).
The rapid decrease of $d_\mathrm{m}$ explains the rapid suppression of $\chi_\mathrm{sc}$ while approaching the topological phase transition at $h\simeq1.305$ in Fig.~\ref{f4}(c).

\section{\label{summary}Concluding remarks}
In conclusion, we have investigated the magnetic HLs in the effective spin model with long-range interactions reflecting the itinerant nature of electrons.
We found that both $4Q$- and $3Q$-HLs are stabilized even at zero magnetic field by the synergy between the DM-type interactions from the spin-orbit coupling and the multiple-spin interactions from the spin-charge coupling.
The results are in stark contrast to the previous studies for the localized spin models with short-range interactions, where the HLs are stable only in a field. 
Furthermore, our HLs may have much shorter periods compared to the previous ones; the periods in our HLs are dictated by nesting properties of the Fermi surface, whereas those in the previous studies are given by the competition between the ferromagnetic exchange interaction and the DM interaction. 
We also clarified the effect of an external magnetic field on the HLs. 
We showed that both $4Q$ and $3Q$ cases exhibit a variety of successive phase transitions depending on the field direction, including the transitions to $2Q$ and $1Q$ states. 
Interestingly, among them, we found several topological phase transitions where the number of monopoles and anti-monopoles changes by the pair annihilation. 
We explicitly showed how the pair annihilation occurs by tracing the real-space positions of monopoles and anti-monopoles on the discrete lattice.

As mentioned in the introduction, $3Q$- and $4Q$-HLs were recently discovered in MnSi$_{1-x}$Ge$_x$~\cite{Tanigaki2015, Kanazawa2017, Fujishiro2019}. 
They are stable even in the absence of the magnetic field and have much shorter periods compared to the conventional skyrmion lattices, for instance, in MnSi, and evaded the understanding from the conventional spin models with short-range two-spin interactions. 
A scenario was recently proposed based on short-range four-spin and six-spin interactions including the scalar spin chirality~\cite{Grytsiuk2019}. 
Our finding suggests another scenario by emphasizing the important role of itinerant nature of electrons.
To test our scenario, it is necessary to clarify the electronic structure in the real compounds, e.g., by the angle-resolved photoemission spectroscopy and the de Haas-van Alphen effect. 
First-principles calculations would also be helpful, while it is not straightforward to precisely predict the relevant wave numbers in the complicated multiorbital systems with electron correlations, in particular, chemically doped materials like MnSi$_{1-x}$Ge$_x$. 
It would also be interesting to test our scenario for the short-period skyrmion lattice recently discovered in EuPtSi~\cite{Kakihana2018,Kaneko2019,Takeuchi2019}.
We note that a similar scenario (without the DM-type interaction) was recently discussed for the swirling spin textures in a centrosymmetric triangular magnet Gd$_2$PdSi$_3$~\cite{Kurumaji2019}.

On the other hand, in the magnetic field, our results suggest that the $4Q$ and $3Q$ states exhibit a nonzero topological Hall effect through the nonzero scalar spin chirality $\chi_\mathrm{sc}$.
Our results also indicate that $\chi_{\rm sc}$ changes drastically corresponding to the modulation of the magnetic textures including the topological transitions by pair annihilations of monopoles and anti-monopoles.
Experimentally, interesting behaviors were observed in a wide range of field and temperature, even with the sign change of the topological Hall resistivity~\cite{Fujishiro2019}. 
Assuming our scenario based on the itinerant nature of electrons, it will be important to take into account the realistic electronic band structures in the magnetic field for detailed comparison between theory and experiment.
In particular, it is worth studying how the modulations of the Fermi surfaces and corresponding $\mathbf{Q}_\eta$ modify the phase diagrams in the magnetic field.
Moreover, thermal fluctuations might also play an important role.
We leave the finite-temperature study as a future work, as it requires sophisticated Monte Carlo simulations beyond the simulated annealing to resolve competing phases.

\begin{acknowledgments}
We would like to thank N.~Kanazawa and K.~Shimizu for fruitful discussions.
This research was supported by Grants-in-Aid for Scientific Research under Grants No.~JP19H05825 and No.~JP18K13488, JST CREST (JP- MJCR18T2), and the Chirality Research Center in Hiroshima University and JSPS Core-to-Core Program, Advanced Research Networks.
S. O. was supported by JSPS through the research fellowship for young scientists.
\end{acknowledgments}

\appendix

\section{\label{appendix:A}Difference among the 4\boldmath{$Q$} states}

\begin{figure}[b]
\centering
\includegraphics[width=\columnwidth,clip]{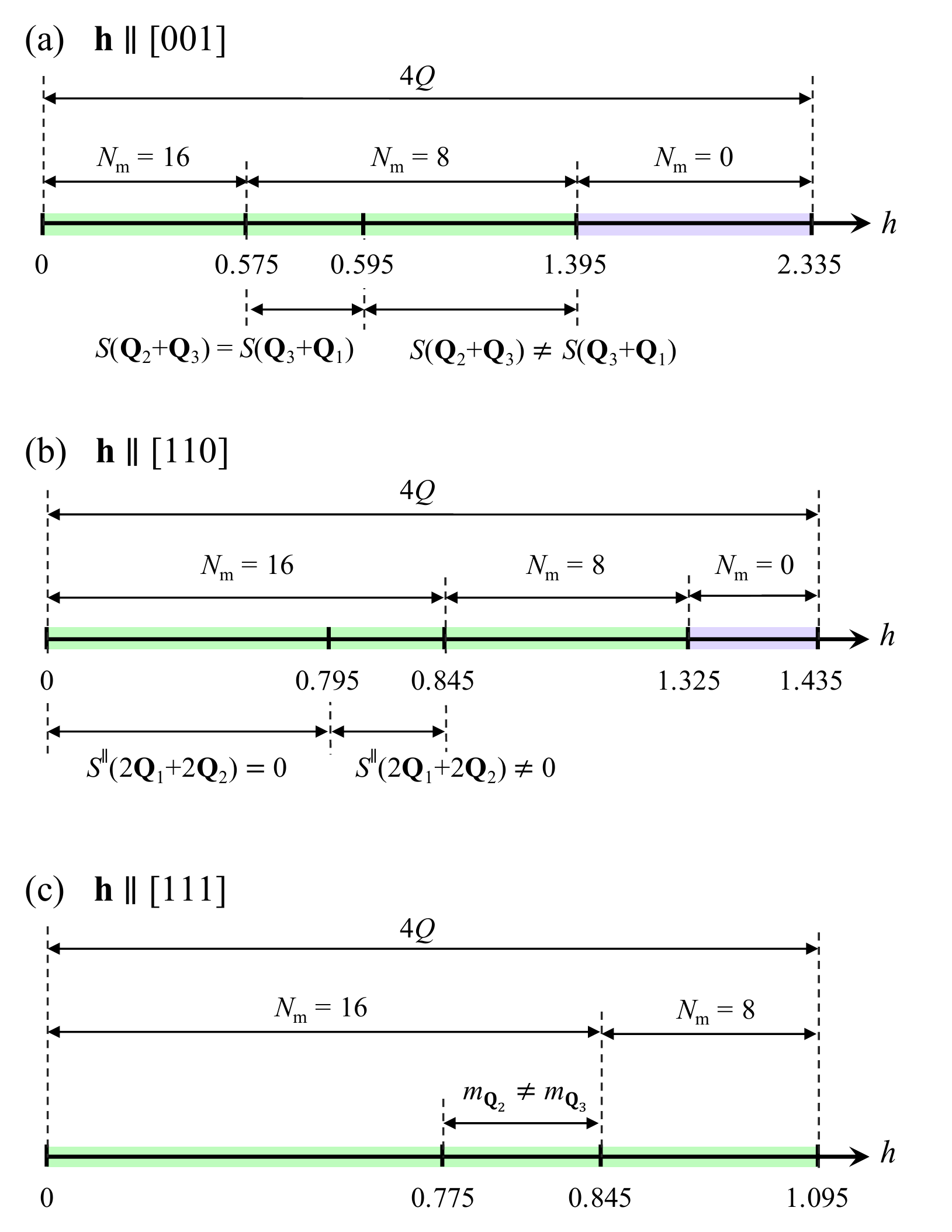}
\caption{
Schematics for the differences among the 4$Q$ phases in the (a) [001], (b) [110], and (c) [111] field corresponding to Figs.~\ref{f3}(a), \ref{f3}(b), and \ref{f3}(c), respectively.
}
\label{fA1}
\end{figure}

In this section, we discuss the difference among the $4Q$ states found in Sec.~\ref{result:4Q}.
In the case of the [001] field in Fig.~\ref{f3}(a), we found four $4Q$ states below $h\simeq 2.335$, all of which have the equal amplitudes of the four $m_{\mathbf{Q}_\eta}$.
Two of them are distinguished by the number of monopoles and anti-monopoles, $N_{\rm m}$: 
the $4Q$-HL state with $N_{\rm m}=16$ below $h\simeq 0.575$ and the $4Q$ state with $N_{\rm m}=0$ above $h\simeq 1.395$. 
The rest two have the same $N_{\rm m}=8$, but we find that they show different values in higher harmonics in the spin structure factor $S(\mathbf{q})$ in Eq.~(\ref{eq:ss_factor}):
By calculating $S(\bold{Q}_\eta+\bold{Q}_{\eta'})$ where $\eta=1,2,3,4$ and $\eta\neq\eta'$, we find that 
$S(\bold{Q}_2+\bold{Q}_3)$ and $S(\bold{Q}_3+\bold{Q}_1)$ have the equal amplitudes for $0.575\lesssim h \lesssim0.595$, but $S(\bold{Q}_2+\bold{Q}_3)\neq S(\bold{Q}_3+\bold{Q}_1)$ for $0.595\lesssim h \lesssim1.395$.
The differences among the four $4Q$ states are summarized in Fig.~\ref{fA1}(a).

In the case of the [110] field in Fig.~\ref{f3}(b), we also find four $4Q$ states below $h\simeq 1.435$ that share the same symmetry in terms of $m_{\mathbf{Q}_\eta}$.
In this case again, $N_{\rm m}$ distinguishes two of them: 
the $4Q$-HL state with $N_{\rm m}=8$ for $0.845\lesssim h \lesssim1.325$ and the $4Q$ state with $N_{\rm m}=0$ for $1.325\lesssim h \lesssim1.435$. 
In order to distinguish the rest two for $h\lesssim0.845$, we calculate the spin structure factor with the spin component parallel to the [110] field defined by  
\begin{align}
S^\parallel(\mathbf{q})=\frac{1}{2}\{S^{xx}(\mathbf{q})+S^{yy}(\mathbf{q})\}+S^{xy}(\mathbf{q}),
\label{eq:ssf_para_110}
\end{align}
where $S^{\mu\nu}(\mathbf{q})$ is the matrix form of the spin structure factor defined by
\begin{align}
S^{\mu\nu}(\mathbf{q})=\frac{1}{N}\sum_{l,l'}S^\mu_{\mathbf{r}_l}S^\nu_{\mathbf{r}_{l'}}e^{i\mathbf{q}\cdot(\mathbf{r}_{l}-\mathbf{r}_{l'})}.
\label{eq:ssf_component}
\end{align}
We find that the higher harmonics along the field direction, $S^\parallel(2\mathbf{Q}_1+2\mathbf{Q}_2)$, is nonzero in the $4Q$ state for $0.795\lesssim h \lesssim0.845$, while it vanishes for $h \lesssim0.795$.
The differences among the four $4Q$ states are summarized in Fig.~\ref{fA1}(b).

Finally, in the case of the [111] field in Fig.~\ref{f3}(c), we find three $4Q$ states.
In this case, $m_{\mathbf{Q}_\eta}$ distinguishes the intermediate phase for $0.845\lesssim h \lesssim1.095$, as mentioned in Sec.~\ref{result:4Q}. 
The rest two can be distinguished by $N_{\rm m}$. 
See Fig.~\ref{fA1}(c).

\section{\label{appendix:B}Difference among the 3\boldmath{$Q$} states}

\begin{figure}[t]
\centering
\includegraphics[width=\columnwidth,clip]{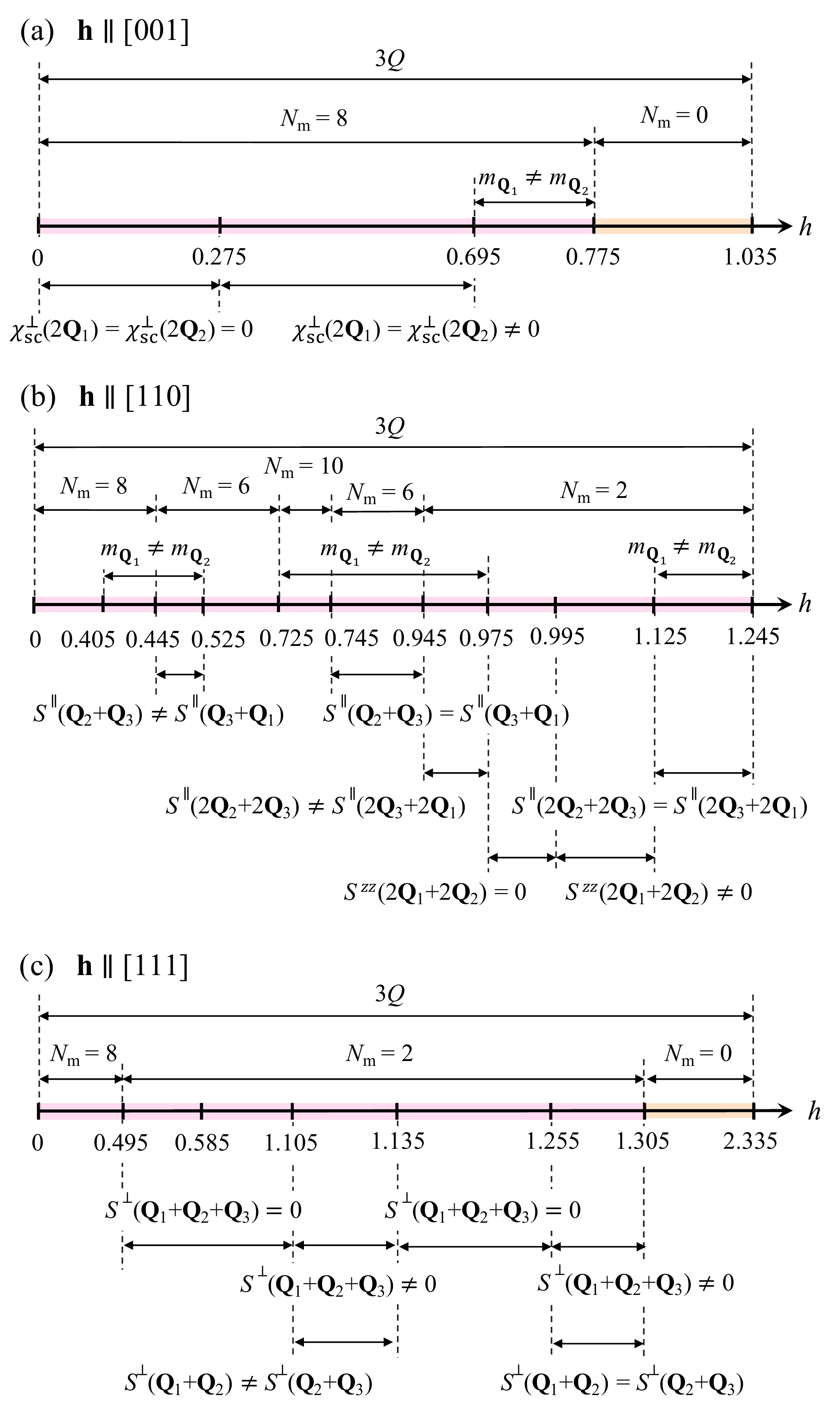}
\caption{
Schematics of the $3Q$ states in the (a) [001], (b) [110], and (c) [111] field corresponding to Figs.~\ref{f4}(a), \ref{f4}(b), and \ref{f4}(c), respectively.
}
\label{fB1}
\end{figure}

In this section, we discuss the difference among the $3Q$ states found in Sec.~\ref{result:3Q}.
In the case of the [001] field in Fig.~\ref{f4}(a), we find four $3Q$ states.
One of them for $0.775\lesssim h \lesssim 1.035$ is distinguished from the others by $N_{\rm m}$: the $3Q$ state with $N_{\rm m}=0$ above $h\simeq 0.775$.
In addition, as mentioned in Sec.~\ref{result:3Q}, $m_{\mathbf{Q}_\eta}$ distinguishes the $3Q$-HL state for $0.695\lesssim h \lesssim0.775$. 
We find a difference between the rest two in higher harmonics in the structure factor of the scalar spin chirality in Eq.~(\ref{eq:scalar_chi_local}).
Specifically, we calculate the component perpendicular to the [001] field defined by
\begin{align}
\chi_\mathrm{sc}^\perp(\mathbf{q})=\frac{1}{2}\{\chi_\mathrm{sc}^{xx}(\mathbf{q})+\chi_\mathrm{sc}^{yy}(\mathbf{q})\}+\chi_\mathrm{sc}^{xy}(\mathbf{q}),
\label{eq:csf_para_110}
\end{align}
where $\chi_\mathrm{sc}^{\mu\nu}(\mathbf{q})$ is the matrix form of the structure factor defined by 
\begin{align}
\chi_\mathrm{sc}^{\mu\nu}(\mathbf{q})=\frac{1}{N}\sum_{l,l'}\chi_\mathrm{sc}^\mu(\mathbf{r}_l)\chi_\mathrm{sc}^\nu(\mathbf{r}_{l'})e^{i\mathbf{q}\cdot(\mathbf{r}_{l}-\mathbf{r}_{l'})}.
\label{eq:csf_component}
\end{align}
We find that $\chi_\mathrm{sc}^\perp(2\mathbf{Q}_1)$ and $\chi_\mathrm{sc}^\perp(2\mathbf{Q}_2)$ are nonzero and have the equal amplitudes in the $3Q$-HL state for $0.275\lesssim h \lesssim0.695$, but vanish below $h\simeq 0.275$.
The differences among the four $3Q$ states are summarized in Fig.~\ref{fB1}(a).

Next, in the case of the [110] field in Fig.~\ref{f4}(b), we found ten $3Q$ states below $h\simeq 1.245$ with a variety of $N_{\rm m}$.
The $3Q$-HL state with $N_{\rm m}=10$ for $0.725 \lesssim h \lesssim 0.745$ is distinguished from others, but $N_{\rm m}=8$ for the two states below $h\simeq 0.445$, $N_{\rm m}=6$ for the three states for $0.445\lesssim h \lesssim0.725$ and $0.745\lesssim h \lesssim0.945$, and $N_{\rm m}=2$ for the rest four for $0.945\lesssim h \lesssim1.245$.
As mentioned in Sec.~\ref{result:3Q}, the two with $N_{\rm m}=8$ below $h\simeq 0.445$ and the two with $N_{\rm m}=6$ for $0.445\lesssim h \lesssim0.725$ are distinguished by $m_{\mathbf{Q}_\eta}$.
Similarly, the two with $N_{\rm m}=2$ for $0.945\lesssim h \lesssim 0.975$ and $1.125\lesssim h \lesssim 1.245$ are distinguished from other two by $m_{\mathbf{Q}_\eta}$.
See Fig.~\ref{fB1}(b).

In order to distinguish the rest, we calculate higher harmonics in the spin structure factor similar to the $4Q$ case in Appendix~\ref{appendix:A}.
For the $3Q$-HL states with $N_{\rm m}=6$, $S^\parallel(\bold{Q}_2+\bold{Q}_3)$ and $S^\parallel(\bold{Q}_3+\bold{Q}_1)$ have the equal amplitudes for $0.745\lesssim h \lesssim0.945$, but do not for $0.445\lesssim h \lesssim0.525$.
Meanwhile, for the two states with $N_{\rm m}=2$ and $m_{\mathbf{Q}_1}\neq m_{\mathbf{Q}_2}$, $S^\parallel(2\bold{Q}_2+2\bold{Q}_3) = S^\parallel(2\bold{Q}_3+2\bold{Q}_1)$ for $1.125\lesssim h \lesssim1.245$ while $S^\parallel(2\bold{Q}_2+2\bold{Q}_3) \neq S^\parallel(2\bold{Q}_3+2\bold{Q}_1)$ for $0.945\lesssim h \lesssim0.975$.
Furthermore, for the two states with $N_{\rm m}=2$ and $m_{\mathbf{Q}_1}=m_{\mathbf{Q}_2}$ the component perpendicular to the [110] field, $S^{zz}(2\bold{Q}_1+2\bold{Q}_2)$, is zero for $0.975\lesssim h \lesssim0.995$, but  nonzero for $0.995\lesssim h \lesssim1.125$.
All the differences among the ten $3Q$ states are summarized in Fig.~\ref{fB1}(b).

Finally, in the case of the [111] field in Fig.~\ref{f4}(c), we found seven $3Q$ states below $h\simeq 2.335$, all of which have the equal amplitudes of the three $m_{\mathbf{Q}_\eta}$.
Two of them are distinguished by $N_{\rm m}$: the $3Q$-HL state with $N_{\rm m}=8$ below $h\simeq 0.495$ and the $3Q$ state with $N_{\rm m}=0$ above $h\simeq 1.305$. 
The rest five have the same $N_{\rm m}=2$, but two of them for $1.105\lesssim h \lesssim1.135$  and $1.255\lesssim h \lesssim1.305$ show nonzero values in higher harmonics in the spin structure factor $S^\perp(\mathbf{Q}_1+\mathbf{Q}_2+\mathbf{Q}_3)$, which is the component perpendicular to the [111] field given by 
\begin{align}
S^\perp(\mathbf{q})=\frac{2}{3}\{S(\mathbf{q})-S^{xy}(\mathbf{q})-S^{yz}(\mathbf{q})-S^{zx}(\mathbf{q})\}.
\label{eq:ssf_perp_111}
\end{align}
Furthermore, $S^\perp(\mathbf{Q}_1+\mathbf{Q}_2) = S^\perp(\mathbf{Q}_2+\mathbf{Q}_3)$ for $1.105\lesssim h \lesssim1.135$, but  $S^\perp(\mathbf{Q}_1+\mathbf{Q}_2) \neq S^\perp(\mathbf{Q}_2+\mathbf{Q}_3)$ for $1.255\lesssim h \lesssim1.305$.
The rest three $3Q$-HLs (for $0.495\lesssim h \lesssim0.585$, $0.585\lesssim h \lesssim1.105$, and $1.135\lesssim h \lesssim1.255$) with $N_{\rm m}=2$ cannot be distinguished within the present analyses although further higher harmonics may tell the difference. 
See Fig.~\ref{fB1}(c).

\bibliography{ref.bib}

\providecommand{\noopsort}[1]{}\providecommand{\singleletter}[1]{#1}%
\begin{thebibliography}{45}%
\makeatletter
\providecommand \@ifxundefined [1]{%
 \@ifx{#1\undefined}
}%
\providecommand \@ifnum [1]{%
 \ifnum #1\expandafter \@firstoftwo
 \else \expandafter \@secondoftwo
 \fi
}%
\providecommand \@ifx [1]{%
 \ifx #1\expandafter \@firstoftwo
 \else \expandafter \@secondoftwo
 \fi
}%
\providecommand \natexlab [1]{#1}%
\providecommand \enquote  [1]{``#1''}%
\providecommand \bibnamefont  [1]{#1}%
\providecommand \bibfnamefont [1]{#1}%
\providecommand \citenamefont [1]{#1}%
\providecommand \href@noop [0]{\@secondoftwo}%
\providecommand \href [0]{\begingroup \@sanitize@url \@href}%
\providecommand \@href[1]{\@@startlink{#1}\@@href}%
\providecommand \@@href[1]{\endgroup#1\@@endlink}%
\providecommand \@sanitize@url [0]{\catcode `\\12\catcode `\$12\catcode
  `\&12\catcode `\#12\catcode `\^12\catcode `\_12\catcode `\%12\relax}%
\providecommand \@@startlink[1]{}%
\providecommand \@@endlink[0]{}%
\providecommand \url  [0]{\begingroup\@sanitize@url \@url }%
\providecommand \@url [1]{\endgroup\@href {#1}{\urlprefix }}%
\providecommand \urlprefix  [0]{URL }%
\providecommand \Eprint [0]{\href }%
\providecommand \doibase [0]{http://dx.doi.org/}%
\providecommand \selectlanguage [0]{\@gobble}%
\providecommand \bibinfo  [0]{\@secondoftwo}%
\providecommand \bibfield  [0]{\@secondoftwo}%
\providecommand \translation [1]{[#1]}%
\providecommand \BibitemOpen [0]{}%
\providecommand \bibitemStop [0]{}%
\providecommand \bibitemNoStop [0]{.\EOS\space}%
\providecommand \EOS [0]{\spacefactor3000\relax}%
\providecommand \BibitemShut  [1]{\csname bibitem#1\endcsname}%
\let\auto@bib@innerbib\@empty
%</preamble>
\bibitem [{\citenamefont {Nagaosa}\ and\ \citenamefont
  {Tokura}(2013)}]{Nagaosa2013}%
  \BibitemOpen
  \bibfield  {author} {\bibinfo {author} {\bibfnamefont {N.}~\bibnamefont
  {Nagaosa}}\ and\ \bibinfo {author} {\bibfnamefont {Y.}~\bibnamefont
  {Tokura}},\ }\bibfield  {title} {\enquote {\bibinfo {title} {Topological
  properties and dynamics of magnetic skyrmions},}\ }\href@noop {} {\bibfield
  {journal} {\bibinfo  {journal} {Nat. Nano.}\ }\textbf {\bibinfo {volume}
  {8}},\ \bibinfo {pages} {899} (\bibinfo {year} {2013})}\BibitemShut {NoStop}%
\bibitem [{\citenamefont {Togawa}\ \emph {et~al.}(2016)\citenamefont {Togawa},
  \citenamefont {Kousaka}, \citenamefont {Inoue},\ and\ \citenamefont
  {Kishine}}]{Togawa2016}%
  \BibitemOpen
  \bibfield  {author} {\bibinfo {author} {\bibfnamefont {Y.}~\bibnamefont
  {Togawa}}, \bibinfo {author} {\bibfnamefont {Y.}~\bibnamefont {Kousaka}},
  \bibinfo {author} {\bibfnamefont {K.}~\bibnamefont {Inoue}}, \ and\ \bibinfo
  {author} {\bibfnamefont {J.}~\bibnamefont {Kishine}},\ }\bibfield  {title}
  {\enquote {\bibinfo {title} {{S}ymmetry, {S}tructure, and {D}ynamics of
  {M}onoaxial {C}hiral {M}agnets},}\ }\href@noop {} {\bibfield  {journal}
  {\bibinfo  {journal} {J. Phys. Soc. Japan}\ }\textbf {\bibinfo {volume}
  {85}},\ \bibinfo {pages} {112001} (\bibinfo {year} {2016})}\BibitemShut
  {NoStop}%
\bibitem [{\citenamefont {Tokura}\ and\ \citenamefont
  {Seki}(2010)}]{Tokura2010}%
  \BibitemOpen
  \bibfield  {author} {\bibinfo {author} {\bibfnamefont {Y.}~\bibnamefont
  {Tokura}}\ and\ \bibinfo {author} {\bibfnamefont {S.}~\bibnamefont {Seki}},\
  }\bibfield  {title} {\enquote {\bibinfo {title} {Multiferroics with {S}piral
  {S}pin {O}rders},}\ }\href@noop {} {\bibfield  {journal} {\bibinfo  {journal}
  {Adv. Mater.}\ }\textbf {\bibinfo {volume} {22}},\ \bibinfo {pages} {1554}
  (\bibinfo {year} {2010})}\BibitemShut {NoStop}%
\bibitem [{\citenamefont {Mochizuki}\ and\ \citenamefont
  {Seki}(2015)}]{Mochizuki2015}%
  \BibitemOpen
  \bibfield  {author} {\bibinfo {author} {\bibfnamefont {M.}~\bibnamefont
  {Mochizuki}}\ and\ \bibinfo {author} {\bibfnamefont {S.}~\bibnamefont
  {Seki}},\ }\bibfield  {title} {\enquote {\bibinfo {title} {Dynamical
  magnetoelectric phenomena of multiferroic skyrmions},}\ }\href@noop {}
  {\bibfield  {journal} {\bibinfo  {journal} {J. Phys.: Condens. Matter}\
  }\textbf {\bibinfo {volume} {27}},\ \bibinfo {pages} {503001} (\bibinfo
  {year} {2015})}\BibitemShut {NoStop}%
\bibitem [{\citenamefont {Tokura}\ and\ \citenamefont
  {Nagaosa}(2018)}]{Tokura2018}%
  \BibitemOpen
  \bibfield  {author} {\bibinfo {author} {\bibfnamefont {Y.}~\bibnamefont
  {Tokura}}\ and\ \bibinfo {author} {\bibfnamefont {N.}~\bibnamefont
  {Nagaosa}},\ }\bibfield  {title} {\enquote {\bibinfo {title} {Nonreciprocal
  responses from non-centrosymmetric quantum materials},}\ }\href@noop {}
  {\bibfield  {journal} {\bibinfo  {journal} {Nat. Commun.}\ }\textbf {\bibinfo
  {volume} {9}},\ \bibinfo {pages} {3740} (\bibinfo {year} {2018})}\BibitemShut
  {NoStop}%
\bibitem [{\citenamefont {Tanigaki}\ \emph {et~al.}(2015)\citenamefont
  {Tanigaki}, \citenamefont {Shibata}, \citenamefont {Kanazawa}, \citenamefont
  {Yu}, \citenamefont {Onose}, \citenamefont {Park}, \citenamefont {Shindo},\
  and\ \citenamefont {Tokura}}]{Tanigaki2015}%
  \BibitemOpen
  \bibfield  {author} {\bibinfo {author} {\bibfnamefont {T.}~\bibnamefont
  {Tanigaki}}, \bibinfo {author} {\bibfnamefont {K.}~\bibnamefont {Shibata}},
  \bibinfo {author} {\bibfnamefont {N.}~\bibnamefont {Kanazawa}}, \bibinfo
  {author} {\bibfnamefont {X.}~\bibnamefont {Yu}}, \bibinfo {author}
  {\bibfnamefont {Y.}~\bibnamefont {Onose}}, \bibinfo {author} {\bibfnamefont
  {H.~S.}\ \bibnamefont {Park}}, \bibinfo {author} {\bibfnamefont
  {D.}~\bibnamefont {Shindo}}, \ and\ \bibinfo {author} {\bibfnamefont
  {Y.}~\bibnamefont {Tokura}},\ }\bibfield  {title} {\enquote {\bibinfo {title}
  {{R}eal-{S}pace {O}bservation of {S}hort-{P}eriod {C}ubic {L}attice of
  {S}kyrmions in {M}n{G}e},}\ }\href@noop {} {\bibfield  {journal} {\bibinfo
  {journal} {Nano Lett.}\ }\textbf {\bibinfo {volume} {15}},\ \bibinfo {pages}
  {5438} (\bibinfo {year} {2015})}\BibitemShut {NoStop}%
\bibitem [{\citenamefont {Kanazawa}\ \emph {et~al.}(2017)\citenamefont
  {Kanazawa}, \citenamefont {Seki},\ and\ \citenamefont
  {Tokura}}]{Kanazawa2017}%
  \BibitemOpen
  \bibfield  {author} {\bibinfo {author} {\bibfnamefont {N.}~\bibnamefont
  {Kanazawa}}, \bibinfo {author} {\bibfnamefont {S.}~\bibnamefont {Seki}}, \
  and\ \bibinfo {author} {\bibfnamefont {Y.}~\bibnamefont {Tokura}},\
  }\bibfield  {title} {\enquote {\bibinfo {title} {{N}oncentrosymmetric
  {M}agnets {H}osting {M}agnetic {S}kyrmions},}\ }\href@noop {} {\bibfield
  {journal} {\bibinfo  {journal} {Adv. Mater.}\ }\textbf {\bibinfo {volume}
  {29}},\ \bibinfo {pages} {1603227} (\bibinfo {year} {2017})}\BibitemShut
  {NoStop}%
\bibitem [{\citenamefont {Kanazawa}\ \emph {et~al.}(2012)\citenamefont
  {Kanazawa}, \citenamefont {Kim}, \citenamefont {Inosov}, \citenamefont
  {White}, \citenamefont {Egetenmeyer}, \citenamefont {Gavilano}, \citenamefont
  {Ishiwata}, \citenamefont {Onose}, \citenamefont {Arima}, \citenamefont
  {Keimer},\ and\ \citenamefont {Tokura}}]{Kanazawa2012}%
  \BibitemOpen
  \bibfield  {author} {\bibinfo {author} {\bibfnamefont {N.}~\bibnamefont
  {Kanazawa}}, \bibinfo {author} {\bibfnamefont {J.-H.}\ \bibnamefont {Kim}},
  \bibinfo {author} {\bibfnamefont {D.~S.}\ \bibnamefont {Inosov}}, \bibinfo
  {author} {\bibfnamefont {J.~S.}\ \bibnamefont {White}}, \bibinfo {author}
  {\bibfnamefont {N.}~\bibnamefont {Egetenmeyer}}, \bibinfo {author}
  {\bibfnamefont {J.~L.}\ \bibnamefont {Gavilano}}, \bibinfo {author}
  {\bibfnamefont {S.}~\bibnamefont {Ishiwata}}, \bibinfo {author}
  {\bibfnamefont {Y.}~\bibnamefont {Onose}}, \bibinfo {author} {\bibfnamefont
  {T.}~\bibnamefont {Arima}}, \bibinfo {author} {\bibfnamefont
  {B.}~\bibnamefont {Keimer}}, \ and\ \bibinfo {author} {\bibfnamefont
  {Y.}~\bibnamefont {Tokura}},\ }\bibfield  {title} {\enquote {\bibinfo {title}
  {Possible skyrmion-lattice ground state in the ${B}20$ chiral-lattice magnet
  {M}n{G}e as seen via small-angle neutron scattering},}\ }\href {\doibase
  10.1103/PhysRevB.86.134425} {\bibfield  {journal} {\bibinfo  {journal} {Phys.
  Rev. B}\ }\textbf {\bibinfo {volume} {86}},\ \bibinfo {pages} {134425}
  (\bibinfo {year} {2012})}\BibitemShut {NoStop}%
\bibitem [{\citenamefont {Kanazawa}\ \emph {et~al.}(2016)\citenamefont
  {Kanazawa}, \citenamefont {Nii}, \citenamefont {Zhang}, \citenamefont
  {Mishchenko}, \citenamefont {Filippis}, \citenamefont {Kagawa}, \citenamefont
  {Iwasa}, \citenamefont {Nagaosa},\ and\ \citenamefont
  {Tokura}}]{Kanazawa2016}%
  \BibitemOpen
  \bibfield  {author} {\bibinfo {author} {\bibfnamefont {N.}~\bibnamefont
  {Kanazawa}}, \bibinfo {author} {\bibfnamefont {Y.}~\bibnamefont {Nii}},
  \bibinfo {author} {\bibfnamefont {X.~X.}\ \bibnamefont {Zhang}}, \bibinfo
  {author} {\bibfnamefont {A.~S.}\ \bibnamefont {Mishchenko}}, \bibinfo
  {author} {\bibfnamefont {G.~De}\ \bibnamefont {Filippis}}, \bibinfo {author}
  {\bibfnamefont {F.}~\bibnamefont {Kagawa}}, \bibinfo {author} {\bibfnamefont
  {Y.}~\bibnamefont {Iwasa}}, \bibinfo {author} {\bibfnamefont
  {N.}~\bibnamefont {Nagaosa}}, \ and\ \bibinfo {author} {\bibfnamefont
  {Y.}~\bibnamefont {Tokura}},\ }\bibfield  {title} {\enquote {\bibinfo {title}
  {Critical phenomena of emergent magnetic monopoles in a chiral magnet},}\
  }\href@noop {} {\bibfield  {journal} {\bibinfo  {journal} {Nat. Commun.}\
  }\textbf {\bibinfo {volume} {7}},\ \bibinfo {pages} {11622} (\bibinfo {year}
  {2016})}\BibitemShut {NoStop}%
\bibitem [{\citenamefont {Zhang}\ \emph {et~al.}(2016)\citenamefont {Zhang},
  \citenamefont {Mishchenko}, \citenamefont {De~Filippis},\ and\ \citenamefont
  {Nagaosa}}]{Zhang2016}%
  \BibitemOpen
  \bibfield  {author} {\bibinfo {author} {\bibfnamefont {X.-X.}\ \bibnamefont
  {Zhang}}, \bibinfo {author} {\bibfnamefont {A.~S.}\ \bibnamefont
  {Mishchenko}}, \bibinfo {author} {\bibfnamefont {G.}~\bibnamefont
  {De~Filippis}}, \ and\ \bibinfo {author} {\bibfnamefont {N.}~\bibnamefont
  {Nagaosa}},\ }\bibfield  {title} {\enquote {\bibinfo {title} {Electric
  transport in three-dimensional skyrmion/monopole crystal},}\ }\href {\doibase
  10.1103/PhysRevB.94.174428} {\bibfield  {journal} {\bibinfo  {journal} {Phys.
  Rev. B}\ }\textbf {\bibinfo {volume} {94}},\ \bibinfo {pages} {174428}
  (\bibinfo {year} {2016})}\BibitemShut {NoStop}%
\bibitem [{\citenamefont {Kanazawa}\ \emph {et~al.}(2011)\citenamefont
  {Kanazawa}, \citenamefont {Onose}, \citenamefont {Arima}, \citenamefont
  {Okuyama}, \citenamefont {Ohoyama}, \citenamefont {Wakimoto}, \citenamefont
  {Kakurai}, \citenamefont {Ishiwata},\ and\ \citenamefont
  {Tokura}}]{Kanazawa2011}%
  \BibitemOpen
  \bibfield  {author} {\bibinfo {author} {\bibfnamefont {N.}~\bibnamefont
  {Kanazawa}}, \bibinfo {author} {\bibfnamefont {Y.}~\bibnamefont {Onose}},
  \bibinfo {author} {\bibfnamefont {T.}~\bibnamefont {Arima}}, \bibinfo
  {author} {\bibfnamefont {D.}~\bibnamefont {Okuyama}}, \bibinfo {author}
  {\bibfnamefont {K.}~\bibnamefont {Ohoyama}}, \bibinfo {author} {\bibfnamefont
  {S.}~\bibnamefont {Wakimoto}}, \bibinfo {author} {\bibfnamefont
  {K.}~\bibnamefont {Kakurai}}, \bibinfo {author} {\bibfnamefont
  {S.}~\bibnamefont {Ishiwata}}, \ and\ \bibinfo {author} {\bibfnamefont
  {Y.}~\bibnamefont {Tokura}},\ }\bibfield  {title} {\enquote {\bibinfo {title}
  {Large {T}opological {H}all {E}ffect in a {S}hort-{P}eriod {H}elimagnet
  {M}n{G}e},}\ }\href {\doibase 10.1103/PhysRevLett.106.156603} {\bibfield
  {journal} {\bibinfo  {journal} {Phys. Rev. Lett.}\ }\textbf {\bibinfo
  {volume} {106}},\ \bibinfo {pages} {156603} (\bibinfo {year}
  {2011})}\BibitemShut {NoStop}%
\bibitem [{\citenamefont {Shiomi}\ \emph {et~al.}(2013)\citenamefont {Shiomi},
  \citenamefont {Kanazawa}, \citenamefont {Shibata}, \citenamefont {Onose},\
  and\ \citenamefont {Tokura}}]{Shiomi2013}%
  \BibitemOpen
  \bibfield  {author} {\bibinfo {author} {\bibfnamefont {Y.}~\bibnamefont
  {Shiomi}}, \bibinfo {author} {\bibfnamefont {N.}~\bibnamefont {Kanazawa}},
  \bibinfo {author} {\bibfnamefont {K.}~\bibnamefont {Shibata}}, \bibinfo
  {author} {\bibfnamefont {Y.}~\bibnamefont {Onose}}, \ and\ \bibinfo {author}
  {\bibfnamefont {Y.}~\bibnamefont {Tokura}},\ }\bibfield  {title} {\enquote
  {\bibinfo {title} {{T}opological {N}ernst effect in a three-dimensional
  skyrmion-lattice phase},}\ }\href {\doibase 10.1103/PhysRevB.88.064409}
  {\bibfield  {journal} {\bibinfo  {journal} {Phys. Rev. B}\ }\textbf {\bibinfo
  {volume} {88}},\ \bibinfo {pages} {064409} (\bibinfo {year}
  {2013})}\BibitemShut {NoStop}%
\bibitem [{\citenamefont {Fujishiro}\ \emph {et~al.}(2018)\citenamefont
  {Fujishiro}, \citenamefont {Kanazawa}, \citenamefont {Shimojima},
  \citenamefont {Nakamura}, \citenamefont {Ishizaka}, \citenamefont
  {Koretsune}, \citenamefont {Arita}, \citenamefont {Miyake}, \citenamefont
  {Mitamura}, \citenamefont {Akiba}, \citenamefont {Tokunaga}, \citenamefont
  {Shiogai}, \citenamefont {Kimura}, \citenamefont {Awaji}, \citenamefont
  {Tsukazaki}, \citenamefont {Kikkawa}, \citenamefont {Taguchi},\ and\
  \citenamefont {Tokura}}]{Fujishiro2018}%
  \BibitemOpen
  \bibfield  {author} {\bibinfo {author} {\bibfnamefont {Y.}~\bibnamefont
  {Fujishiro}}, \bibinfo {author} {\bibfnamefont {N.}~\bibnamefont {Kanazawa}},
  \bibinfo {author} {\bibfnamefont {T.}~\bibnamefont {Shimojima}}, \bibinfo
  {author} {\bibfnamefont {A.}~\bibnamefont {Nakamura}}, \bibinfo {author}
  {\bibfnamefont {K.}~\bibnamefont {Ishizaka}}, \bibinfo {author}
  {\bibfnamefont {T.}~\bibnamefont {Koretsune}}, \bibinfo {author}
  {\bibfnamefont {R.}~\bibnamefont {Arita}}, \bibinfo {author} {\bibfnamefont
  {A.}~\bibnamefont {Miyake}}, \bibinfo {author} {\bibfnamefont
  {H.}~\bibnamefont {Mitamura}}, \bibinfo {author} {\bibfnamefont
  {K.}~\bibnamefont {Akiba}}, \bibinfo {author} {\bibfnamefont
  {M.}~\bibnamefont {Tokunaga}}, \bibinfo {author} {\bibfnamefont
  {J.}~\bibnamefont {Shiogai}}, \bibinfo {author} {\bibfnamefont
  {S.}~\bibnamefont {Kimura}}, \bibinfo {author} {\bibfnamefont
  {S.}~\bibnamefont {Awaji}}, \bibinfo {author} {\bibfnamefont
  {A.}~\bibnamefont {Tsukazaki}}, \bibinfo {author} {\bibfnamefont
  {A.}~\bibnamefont {Kikkawa}}, \bibinfo {author} {\bibfnamefont
  {Y.}~\bibnamefont {Taguchi}}, \ and\ \bibinfo {author} {\bibfnamefont
  {Y.}~\bibnamefont {Tokura}},\ }\bibfield  {title} {\enquote {\bibinfo {title}
  {Large magneto-thermopower in {M}n{G}e with topological spin texture},}\
  }\href@noop {} {\bibfield  {journal} {\bibinfo  {journal} {Nat. Commun.}\
  }\textbf {\bibinfo {volume} {9}},\ \bibinfo {pages} {408} (\bibinfo {year}
  {2018})}\BibitemShut {NoStop}%
\bibitem [{\citenamefont {Fujishiro}\ \emph {et~al.}(2019)\citenamefont
  {Fujishiro}, \citenamefont {Kanazawa}, \citenamefont {Nakajima},
  \citenamefont {Yu}, \citenamefont {Ohishi}, \citenamefont {Kawamura},
  \citenamefont {Kakurai}, \citenamefont {Arima}, \citenamefont {Mitamura},
  \citenamefont {Miyake}, \citenamefont {Akiba}, \citenamefont {Tokunaga},
  \citenamefont {Matsuo}, \citenamefont {Kindo}, \citenamefont {Koretsune},
  \citenamefont {Arita},\ and\ \citenamefont {Tokura}}]{Fujishiro2019}%
  \BibitemOpen
  \bibfield  {author} {\bibinfo {author} {\bibfnamefont {Y.}~\bibnamefont
  {Fujishiro}}, \bibinfo {author} {\bibfnamefont {N.}~\bibnamefont {Kanazawa}},
  \bibinfo {author} {\bibfnamefont {T.}~\bibnamefont {Nakajima}}, \bibinfo
  {author} {\bibfnamefont {X.~Z.}\ \bibnamefont {Yu}}, \bibinfo {author}
  {\bibfnamefont {K.}~\bibnamefont {Ohishi}}, \bibinfo {author} {\bibfnamefont
  {Y.}~\bibnamefont {Kawamura}}, \bibinfo {author} {\bibfnamefont
  {K.}~\bibnamefont {Kakurai}}, \bibinfo {author} {\bibfnamefont
  {T.}~\bibnamefont {Arima}}, \bibinfo {author} {\bibfnamefont
  {H.}~\bibnamefont {Mitamura}}, \bibinfo {author} {\bibfnamefont
  {A.}~\bibnamefont {Miyake}}, \bibinfo {author} {\bibfnamefont
  {K.}~\bibnamefont {Akiba}}, \bibinfo {author} {\bibfnamefont
  {M.}~\bibnamefont {Tokunaga}}, \bibinfo {author} {\bibfnamefont
  {A.}~\bibnamefont {Matsuo}}, \bibinfo {author} {\bibfnamefont
  {K.}~\bibnamefont {Kindo}}, \bibinfo {author} {\bibfnamefont
  {T.}~\bibnamefont {Koretsune}}, \bibinfo {author} {\bibfnamefont
  {R.}~\bibnamefont {Arita}}, \ and\ \bibinfo {author} {\bibfnamefont
  {Y.}~\bibnamefont {Tokura}},\ }\bibfield  {title} {\enquote {\bibinfo {title}
  {Topological transitions among skyrmion- and hedgehog-lattice states in cubic
  chiral magnets},}\ }\href@noop {} {\bibfield  {journal} {\bibinfo  {journal}
  {Nat. Commun.}\ }\textbf {\bibinfo {volume} {10}},\ \bibinfo {pages} {1059}
  (\bibinfo {year} {2019})}\BibitemShut {NoStop}%
\bibitem [{\citenamefont {Binz}\ and\ \citenamefont
  {Vishwanath}(2006)}]{Binz2006}%
  \BibitemOpen
  \bibfield  {author} {\bibinfo {author} {\bibfnamefont {B.}~\bibnamefont
  {Binz}}\ and\ \bibinfo {author} {\bibfnamefont {A.}~\bibnamefont
  {Vishwanath}},\ }\bibfield  {title} {\enquote {\bibinfo {title} {Theory of
  helical spin crystals: {P}hases, textures, and properties},}\ }\href
  {\doibase 10.1103/PhysRevB.74.214408} {\bibfield  {journal} {\bibinfo
  {journal} {Phys. Rev. B}\ }\textbf {\bibinfo {volume} {74}},\ \bibinfo
  {pages} {214408} (\bibinfo {year} {2006})}\BibitemShut {NoStop}%
\bibitem [{\citenamefont {Park}\ and\ \citenamefont {Han}(2011)}]{Park2011}%
  \BibitemOpen
  \bibfield  {author} {\bibinfo {author} {\bibfnamefont {J.-H.}\ \bibnamefont
  {Park}}\ and\ \bibinfo {author} {\bibfnamefont {J.~H.}\ \bibnamefont {Han}},\
  }\bibfield  {title} {\enquote {\bibinfo {title} {Zero-temperature phases for
  chiral magnets in three dimensions},}\ }\href {\doibase
  10.1103/PhysRevB.83.184406} {\bibfield  {journal} {\bibinfo  {journal} {Phys.
  Rev. B}\ }\textbf {\bibinfo {volume} {83}},\ \bibinfo {pages} {184406}
  (\bibinfo {year} {2011})}\BibitemShut {NoStop}%
\bibitem [{\citenamefont {Yang}\ \emph {et~al.}(2016)\citenamefont {Yang},
  \citenamefont {Liu},\ and\ \citenamefont {Han}}]{Yang2016}%
  \BibitemOpen
  \bibfield  {author} {\bibinfo {author} {\bibfnamefont {S.-G.}\ \bibnamefont
  {Yang}}, \bibinfo {author} {\bibfnamefont {Y.-H.}\ \bibnamefont {Liu}}, \
  and\ \bibinfo {author} {\bibfnamefont {J.~H.}\ \bibnamefont {Han}},\
  }\bibfield  {title} {\enquote {\bibinfo {title} {Formation of a topological
  monopole lattice and its dynamics in three-dimensional chiral magnets},}\
  }\href {\doibase 10.1103/PhysRevB.94.054420} {\bibfield  {journal} {\bibinfo
  {journal} {Phys. Rev. B}\ }\textbf {\bibinfo {volume} {94}},\ \bibinfo
  {pages} {054420} (\bibinfo {year} {2016})}\BibitemShut {NoStop}%
\bibitem [{\citenamefont {Dzyaloshinsky}(1958)}]{Dzyaloshinskii1958}%
  \BibitemOpen
  \bibfield  {author} {\bibinfo {author} {\bibfnamefont {I.}~\bibnamefont
  {Dzyaloshinsky}},\ }\bibfield  {title} {\enquote {\bibinfo {title} {A
  thermodynamic theory of “weak” ferromagnetism of antiferromagnetics},}\
  }\href@noop {} {\bibfield  {journal} {\bibinfo  {journal} {J. Phys. Chem.
  Solids}\ }\textbf {\bibinfo {volume} {4}},\ \bibinfo {pages} {241} (\bibinfo
  {year} {1958})}\BibitemShut {NoStop}%
\bibitem [{\citenamefont {Moriya}(1960)}]{Moriya1960}%
  \BibitemOpen
  \bibfield  {author} {\bibinfo {author} {\bibfnamefont {T.}~\bibnamefont
  {Moriya}},\ }\bibfield  {title} {\enquote {\bibinfo {title} {{A}nisotropic
  {S}uperexchange {I}nteraction and {W}eak {F}erromagnetism},}\ }\href
  {\doibase 10.1103/PhysRev.120.91} {\bibfield  {journal} {\bibinfo  {journal}
  {Phys. Rev.}\ }\textbf {\bibinfo {volume} {120}},\ \bibinfo {pages} {91--98}
  (\bibinfo {year} {1960})}\BibitemShut {NoStop}%
\bibitem [{\citenamefont {Gayles}\ \emph {et~al.}(2015)\citenamefont {Gayles},
  \citenamefont {Freimuth}, \citenamefont {Schena}, \citenamefont {Lani},
  \citenamefont {Mavropoulos}, \citenamefont {Duine}, \citenamefont {Bl\"ugel},
  \citenamefont {Sinova},\ and\ \citenamefont {Mokrousov}}]{Gayles2015}%
  \BibitemOpen
  \bibfield  {author} {\bibinfo {author} {\bibfnamefont {J.}~\bibnamefont
  {Gayles}}, \bibinfo {author} {\bibfnamefont {F.}~\bibnamefont {Freimuth}},
  \bibinfo {author} {\bibfnamefont {T.}~\bibnamefont {Schena}}, \bibinfo
  {author} {\bibfnamefont {G.}~\bibnamefont {Lani}}, \bibinfo {author}
  {\bibfnamefont {P.}~\bibnamefont {Mavropoulos}}, \bibinfo {author}
  {\bibfnamefont {R.~A.}\ \bibnamefont {Duine}}, \bibinfo {author}
  {\bibfnamefont {S.}~\bibnamefont {Bl\"ugel}}, \bibinfo {author}
  {\bibfnamefont {J.}~\bibnamefont {Sinova}}, \ and\ \bibinfo {author}
  {\bibfnamefont {Y.}~\bibnamefont {Mokrousov}},\ }\bibfield  {title} {\enquote
  {\bibinfo {title} {{D}zyaloshinskii-{M}oriya {I}nteraction and {H}all
  {E}ffects in the {S}kyrmion {P}hase of
  {M}n$_{1\ensuremath{-}x}${F}e$_{x}${G}e},}\ }\href {\doibase
  10.1103/PhysRevLett.115.036602} {\bibfield  {journal} {\bibinfo  {journal}
  {Phys. Rev. Lett.}\ }\textbf {\bibinfo {volume} {115}},\ \bibinfo {pages}
  {036602} (\bibinfo {year} {2015})}\BibitemShut {NoStop}%
\bibitem [{\citenamefont {Koretsune}\ \emph {et~al.}(2015)\citenamefont
  {Koretsune}, \citenamefont {Nagaosa},\ and\ \citenamefont
  {Arita}}]{Koretsune2015}%
  \BibitemOpen
  \bibfield  {author} {\bibinfo {author} {\bibfnamefont {T.}~\bibnamefont
  {Koretsune}}, \bibinfo {author} {\bibfnamefont {N.}~\bibnamefont {Nagaosa}},
  \ and\ \bibinfo {author} {\bibfnamefont {R}~\bibnamefont {Arita}},\
  }\bibfield  {title} {\enquote {\bibinfo {title} {{C}ontrol of
  {D}zyaloshinskii-{M}oriya interaction in
  {M}n$_{1\ensuremath{-}x}${F}e$_{x}${G}e: a first-principles study},}\
  }\href@noop {} {\bibfield  {journal} {\bibinfo  {journal} {Sci. Rep.}\
  }\textbf {\bibinfo {volume} {5}},\ \bibinfo {pages} {13302} (\bibinfo {year}
  {2015})}\BibitemShut {NoStop}%
\bibitem [{\citenamefont {Kikuchi}\ \emph {et~al.}(2016)\citenamefont
  {Kikuchi}, \citenamefont {Koretsune}, \citenamefont {Arita},\ and\
  \citenamefont {Tatara}}]{Kikuchi2016}%
  \BibitemOpen
  \bibfield  {author} {\bibinfo {author} {\bibfnamefont {Toru}\ \bibnamefont
  {Kikuchi}}, \bibinfo {author} {\bibfnamefont {Takashi}\ \bibnamefont
  {Koretsune}}, \bibinfo {author} {\bibfnamefont {Ryotaro}\ \bibnamefont
  {Arita}}, \ and\ \bibinfo {author} {\bibfnamefont {Gen}\ \bibnamefont
  {Tatara}},\ }\bibfield  {title} {\enquote {\bibinfo {title}
  {{D}zyaloshinskii-{M}oriya {I}nteraction as a {C}onsequence of a {D}oppler
  {S}hift due to {S}pin-{O}rbit-{I}nduced {I}ntrinsic {S}pin {C}urrent},}\
  }\href {\doibase 10.1103/PhysRevLett.116.247201} {\bibfield  {journal}
  {\bibinfo  {journal} {Phys. Rev. Lett.}\ }\textbf {\bibinfo {volume} {116}},\
  \bibinfo {pages} {247201} (\bibinfo {year} {2016})}\BibitemShut {NoStop}%
\bibitem [{\citenamefont {Grytsiuk}\ \emph {et~al.}(2019)\citenamefont
  {Grytsiuk}, \citenamefont {Hoffmann}, \citenamefont {Hanke}, \citenamefont
  {Mavropoulos}, \citenamefont {Mokrousov}, \citenamefont {Bihlmayer},\ and\
  \citenamefont {Bl\"ugel}}]{Grytsiuk2019}%
  \BibitemOpen
  \bibfield  {author} {\bibinfo {author} {\bibfnamefont {S.}~\bibnamefont
  {Grytsiuk}}, \bibinfo {author} {\bibfnamefont {M.}~\bibnamefont {Hoffmann}},
  \bibinfo {author} {\bibfnamefont {J.-P.}\ \bibnamefont {Hanke}}, \bibinfo
  {author} {\bibfnamefont {P.}~\bibnamefont {Mavropoulos}}, \bibinfo {author}
  {\bibfnamefont {Y.}~\bibnamefont {Mokrousov}}, \bibinfo {author}
  {\bibfnamefont {G.}~\bibnamefont {Bihlmayer}}, \ and\ \bibinfo {author}
  {\bibfnamefont {S.}~\bibnamefont {Bl\"ugel}},\ }\bibfield  {title} {\enquote
  {\bibinfo {title} {Ab initio analysis of magnetic properties of the prototype
  {B}20 chiral magnet {F}e{G}e},}\ }\href {\doibase
  10.1103/PhysRevB.100.214406} {\bibfield  {journal} {\bibinfo  {journal}
  {Phys. Rev. B}\ }\textbf {\bibinfo {volume} {100}},\ \bibinfo {pages}
  {214406} (\bibinfo {year} {2019})}\BibitemShut {NoStop}%
\bibitem [{\citenamefont {Brinker}\ \emph {et~al.}(2019)\citenamefont
  {Brinker}, \citenamefont {Dias},\ and\ \citenamefont {Lounis}}]{Brinker2019}%
  \BibitemOpen
  \bibfield  {author} {\bibinfo {author} {\bibfnamefont {S.}~\bibnamefont
  {Brinker}}, \bibinfo {author} {\bibfnamefont {M.~dos~S.}\ \bibnamefont
  {Dias}}, \ and\ \bibinfo {author} {\bibfnamefont {S.}~\bibnamefont
  {Lounis}},\ }\bibfield  {title} {\enquote {\bibinfo {title} {The chiral
  biquadratic pair interaction},}\ }\href@noop {} {\bibfield  {journal}
  {\bibinfo  {journal} {New J. Phys.}\ }\textbf {\bibinfo {volume} {21}},\
  \bibinfo {pages} {083015} (\bibinfo {year} {2019})}\BibitemShut {NoStop}%
\bibitem [{\citenamefont {Grytsiuk}\ \emph {et~al.}(2020)\citenamefont
  {Grytsiuk}, \citenamefont {Hanke}, \citenamefont {Hoffmann}, \citenamefont
  {Bouaziz}, \citenamefont {Gomonay}, \citenamefont {Bihlmayer}, \citenamefont
  {Mokrousov},\ and\ \citenamefont {Bl\"ugel}}]{Grytsiuk2020}%
  \BibitemOpen
  \bibfield  {author} {\bibinfo {author} {\bibfnamefont {S.}~\bibnamefont
  {Grytsiuk}}, \bibinfo {author} {\bibfnamefont {J.-P.}\ \bibnamefont {Hanke}},
  \bibinfo {author} {\bibfnamefont {M.}~\bibnamefont {Hoffmann}}, \bibinfo
  {author} {\bibfnamefont {J.}~\bibnamefont {Bouaziz}}, \bibinfo {author}
  {\bibfnamefont {O.}~\bibnamefont {Gomonay}}, \bibinfo {author} {\bibfnamefont
  {G.}~\bibnamefont {Bihlmayer}}, \bibinfo {author} {\bibfnamefont
  {Y.}~\bibnamefont {Mokrousov}}, \ and\ \bibinfo {author} {\bibfnamefont
  {S.}~\bibnamefont {Bl\"ugel}},\ }\bibfield  {title} {\enquote {\bibinfo
  {title} {Topological^^e2^^80^^93chiral magnetic interactions driven by
  emergent orbital magnetism},}\ }\href@noop {} {\bibfield  {journal} {\bibinfo
   {journal} {Nat. Commun.}\ }\textbf {\bibinfo {volume} {11}},\ \bibinfo
  {pages} {511} (\bibinfo {year} {2020})}\BibitemShut {NoStop}%
\bibitem [{\citenamefont {Stewart}(1984)}]{Stewart1984}%
  \BibitemOpen
  \bibfield  {author} {\bibinfo {author} {\bibfnamefont {G.~R.}\ \bibnamefont
  {Stewart}},\ }\bibfield  {title} {\enquote {\bibinfo {title} {Heavy-fermion
  systems},}\ }\href {\doibase 10.1103/RevModPhys.56.755} {\bibfield  {journal}
  {\bibinfo  {journal} {Rev. Mod. Phys.}\ }\textbf {\bibinfo {volume} {56}},\
  \bibinfo {pages} {755--787} (\bibinfo {year} {1984})}\BibitemShut {NoStop}%
\bibitem [{\citenamefont {Gegenwart}\ \emph {et~al.}(2008)\citenamefont
  {Gegenwart}, \citenamefont {Si},\ and\ \citenamefont
  {Steglich}}]{Gegenwart2008}%
  \BibitemOpen
  \bibfield  {author} {\bibinfo {author} {\bibfnamefont {P.}~\bibnamefont
  {Gegenwart}}, \bibinfo {author} {\bibfnamefont {Q.}~\bibnamefont {Si}}, \
  and\ \bibinfo {author} {\bibfnamefont {F.}~\bibnamefont {Steglich}},\
  }\bibfield  {title} {\enquote {\bibinfo {title} {Quantum criticality in
  heavy-fermion metals},}\ }\href@noop {} {\bibfield  {journal} {\bibinfo
  {journal} {Nat. Phys.}\ }\textbf {\bibinfo {volume} {4}},\ \bibinfo {pages}
  {186--197} (\bibinfo {year} {2008})}\BibitemShut {NoStop}%
\bibitem [{\citenamefont {Martin}\ and\ \citenamefont
  {Batista}(2008)}]{Martin2008}%
  \BibitemOpen
  \bibfield  {author} {\bibinfo {author} {\bibfnamefont {I.}~\bibnamefont
  {Martin}}\ and\ \bibinfo {author} {\bibfnamefont {C.~D.}\ \bibnamefont
  {Batista}},\ }\bibfield  {title} {\enquote {\bibinfo {title} {{I}tinerant
  {E}lectron-{D}riven {C}hiral {M}agnetic {O}rdering and {S}pontaneous
  {Q}uantum {H}all {E}ffect in {T}riangular {L}attice {M}odels},}\ }\href
  {\doibase 10.1103/PhysRevLett.101.156402} {\bibfield  {journal} {\bibinfo
  {journal} {Phys. Rev. Lett.}\ }\textbf {\bibinfo {volume} {101}},\ \bibinfo
  {pages} {156402} (\bibinfo {year} {2008})}\BibitemShut {NoStop}%
\bibitem [{\citenamefont {Zener}(1951)}]{Zener1951}%
  \BibitemOpen
  \bibfield  {author} {\bibinfo {author} {\bibfnamefont {C.}~\bibnamefont
  {Zener}},\ }\bibfield  {title} {\enquote {\bibinfo {title} {{I}nteraction
  between the $d$-{S}hells in the {T}ransition {M}etals. {II}. {F}erromagnetic
  {C}ompounds of {M}anganese with {P}erovskite {S}tructure},}\ }\href {\doibase
  10.1103/PhysRev.82.403} {\bibfield  {journal} {\bibinfo  {journal} {Phys.
  Rev.}\ }\textbf {\bibinfo {volume} {82}},\ \bibinfo {pages} {403} (\bibinfo
  {year} {1951})}\BibitemShut {NoStop}%
\bibitem [{\citenamefont {Anderson}\ and\ \citenamefont
  {Hasegawa}(1955)}]{Anderson1955}%
  \BibitemOpen
  \bibfield  {author} {\bibinfo {author} {\bibfnamefont {P.~W.}\ \bibnamefont
  {Anderson}}\ and\ \bibinfo {author} {\bibfnamefont {H.}~\bibnamefont
  {Hasegawa}},\ }\bibfield  {title} {\enquote {\bibinfo {title}
  {{C}onsiderations on {D}ouble {E}xchange},}\ }\href {\doibase
  10.1103/PhysRev.100.675} {\bibfield  {journal} {\bibinfo  {journal} {Phys.
  Rev.}\ }\textbf {\bibinfo {volume} {100}},\ \bibinfo {pages} {675} (\bibinfo
  {year} {1955})}\BibitemShut {NoStop}%
\bibitem [{\citenamefont {Ruderman}\ and\ \citenamefont
  {Kittel}(1954)}]{Ruderman1954}%
  \BibitemOpen
  \bibfield  {author} {\bibinfo {author} {\bibfnamefont {M.~A.}\ \bibnamefont
  {Ruderman}}\ and\ \bibinfo {author} {\bibfnamefont {C.}~\bibnamefont
  {Kittel}},\ }\bibfield  {title} {\enquote {\bibinfo {title} {{I}ndirect
  {E}xchange {C}oupling of {N}uclear {M}agnetic {M}oments by {C}onduction
  {E}lectrons},}\ }\href {\doibase 10.1103/PhysRev.96.99} {\bibfield  {journal}
  {\bibinfo  {journal} {Phys. Rev.}\ }\textbf {\bibinfo {volume} {96}},\
  \bibinfo {pages} {99--102} (\bibinfo {year} {1954})}\BibitemShut {NoStop}%
\bibitem [{\citenamefont {Kasuya}(1956)}]{Kasuya1956}%
  \BibitemOpen
  \bibfield  {author} {\bibinfo {author} {\bibfnamefont {T.}~\bibnamefont
  {Kasuya}},\ }\bibfield  {title} {\enquote {\bibinfo {title} {{A} {T}heory of
  {M}etallic {F}erro- and {A}ntiferromagnetism on {Z}ener's {M}odel},}\
  }\href@noop {} {\bibfield  {journal} {\bibinfo  {journal} {Prog. Theor.
  Phys.}\ }\textbf {\bibinfo {volume} {16}},\ \bibinfo {pages} {45} (\bibinfo
  {year} {1956})}\BibitemShut {NoStop}%
\bibitem [{\citenamefont {Yosida}(1957)}]{Yosida1957}%
  \BibitemOpen
  \bibfield  {author} {\bibinfo {author} {\bibfnamefont {K.}~\bibnamefont
  {Yosida}},\ }\bibfield  {title} {\enquote {\bibinfo {title} {{M}agnetic
  {P}roperties of {C}u-{M}n {A}lloys},}\ }\href {\doibase
  10.1103/PhysRev.106.893} {\bibfield  {journal} {\bibinfo  {journal} {Phys.
  Rev.}\ }\textbf {\bibinfo {volume} {106}},\ \bibinfo {pages} {893--898}
  (\bibinfo {year} {1957})}\BibitemShut {NoStop}%
\bibitem [{\citenamefont {Akagi}\ \emph {et~al.}(2012)\citenamefont {Akagi},
  \citenamefont {Udagawa},\ and\ \citenamefont {Motome}}]{Akagi2012}%
  \BibitemOpen
  \bibfield  {author} {\bibinfo {author} {\bibfnamefont {Y.}~\bibnamefont
  {Akagi}}, \bibinfo {author} {\bibfnamefont {M.}~\bibnamefont {Udagawa}}, \
  and\ \bibinfo {author} {\bibfnamefont {Y.}~\bibnamefont {Motome}},\
  }\bibfield  {title} {\enquote {\bibinfo {title} {{H}idden {M}ultiple-{S}pin
  {I}nteractions as an {O}rigin of {S}pin {S}calar {C}hiral {O}rder in
  {F}rustrated {K}ondo {L}attice {M}odels},}\ }\href {\doibase
  10.1103/PhysRevLett.108.096401} {\bibfield  {journal} {\bibinfo  {journal}
  {Phys. Rev. Lett.}\ }\textbf {\bibinfo {volume} {108}},\ \bibinfo {pages}
  {096401} (\bibinfo {year} {2012})}\BibitemShut {NoStop}%
\bibitem [{\citenamefont {Hayami}\ and\ \citenamefont
  {Motome}(2014)}]{Hayami2014}%
  \BibitemOpen
  \bibfield  {author} {\bibinfo {author} {\bibfnamefont {S.}~\bibnamefont
  {Hayami}}\ and\ \bibinfo {author} {\bibfnamefont {Y.}~\bibnamefont
  {Motome}},\ }\bibfield  {title} {\enquote {\bibinfo {title} {Multiple-${Q}$
  instability by $(d\ensuremath{-}2)$-dimensional connections of {F}ermi
  surfaces},}\ }\href {\doibase 10.1103/PhysRevB.90.060402} {\bibfield
  {journal} {\bibinfo  {journal} {Phys. Rev. B}\ }\textbf {\bibinfo {volume}
  {90}},\ \bibinfo {pages} {060402} (\bibinfo {year} {2014})}\BibitemShut
  {NoStop}%
\bibitem [{\citenamefont {Hayami}\ \emph {et~al.}(2017)\citenamefont {Hayami},
  \citenamefont {Ozawa},\ and\ \citenamefont {Motome}}]{Hayami2017}%
  \BibitemOpen
  \bibfield  {author} {\bibinfo {author} {\bibfnamefont {S.}~\bibnamefont
  {Hayami}}, \bibinfo {author} {\bibfnamefont {R.}~\bibnamefont {Ozawa}}, \
  and\ \bibinfo {author} {\bibfnamefont {Y.}~\bibnamefont {Motome}},\
  }\bibfield  {title} {\enquote {\bibinfo {title} {Effective
  bilinear-biquadratic model for noncoplanar ordering in itinerant magnets},}\
  }\href {\doibase 10.1103/PhysRevB.95.224424} {\bibfield  {journal} {\bibinfo
  {journal} {Phys. Rev. B}\ }\textbf {\bibinfo {volume} {95}},\ \bibinfo
  {pages} {224424} (\bibinfo {year} {2017})}\BibitemShut {NoStop}%
\bibitem [{\citenamefont {Hayami}\ and\ \citenamefont
  {Motome}(2018)}]{Hayami2018}%
  \BibitemOpen
  \bibfield  {author} {\bibinfo {author} {\bibfnamefont {S.}~\bibnamefont
  {Hayami}}\ and\ \bibinfo {author} {\bibfnamefont {Y.}~\bibnamefont
  {Motome}},\ }\bibfield  {title} {\enquote {\bibinfo {title} {N\'eel- and
  {B}loch-{T}ype {M}agnetic {V}ortices in {R}ashba {M}etals},}\ }\href
  {\doibase 10.1103/PhysRevLett.121.137202} {\bibfield  {journal} {\bibinfo
  {journal} {Phys. Rev. Lett.}\ }\textbf {\bibinfo {volume} {121}},\ \bibinfo
  {pages} {137202} (\bibinfo {year} {2018})}\BibitemShut {NoStop}%
\bibitem [{Note1()}]{Note1}%
  \BibitemOpen
  \bibinfo {note} {The perturbation expansion for Eq.~(\ref {eq:KLM}) leads to
  $\protect \mathbf {D}_\eta \parallel \protect \mathbf {Q}_\eta $~\cite
  {Hayami2018}}\BibitemShut {NoStop}%
\bibitem [{\citenamefont {Ozawa}\ \emph {et~al.}(2016)\citenamefont {Ozawa},
  \citenamefont {Hayami}, \citenamefont {Barros}, \citenamefont {Chern},
  \citenamefont {Motome},\ and\ \citenamefont {Batista}}]{Ozawa2016}%
  \BibitemOpen
  \bibfield  {author} {\bibinfo {author} {\bibfnamefont {R.}~\bibnamefont
  {Ozawa}}, \bibinfo {author} {\bibfnamefont {S.}~\bibnamefont {Hayami}},
  \bibinfo {author} {\bibfnamefont {K.}~\bibnamefont {Barros}}, \bibinfo
  {author} {\bibfnamefont {G.-W.}\ \bibnamefont {Chern}}, \bibinfo {author}
  {\bibfnamefont {Y.}~\bibnamefont {Motome}}, \ and\ \bibinfo {author}
  {\bibfnamefont {C.~D.}\ \bibnamefont {Batista}},\ }\bibfield  {title}
  {\enquote {\bibinfo {title} {Vortex {C}rystals with {C}hiral {S}tripes in
  {I}tinerant {M}agnets},}\ }\href@noop {} {\bibfield  {journal} {\bibinfo
  {journal} {J. Phys. Soc. Japan}\ }\textbf {\bibinfo {volume} {85}},\ \bibinfo
  {pages} {103703} (\bibinfo {year} {2016})}\BibitemShut {NoStop}%
\bibitem [{\citenamefont {Okumura}\ \emph {et~al.}()\citenamefont {Okumura},
  \citenamefont {Hayami}, \citenamefont {Kato},\ and\ \citenamefont
  {Motome}}]{Okumura2019JPSCP}%
  \BibitemOpen
  \bibfield  {author} {\bibinfo {author} {\bibfnamefont {S.}~\bibnamefont
  {Okumura}}, \bibinfo {author} {\bibfnamefont {S.}~\bibnamefont {Hayami}},
  \bibinfo {author} {\bibfnamefont {Y.}~\bibnamefont {Kato}}, \ and\ \bibinfo
  {author} {\bibfnamefont {Y.}~\bibnamefont {Motome}},\ }\href@noop {}
  {\enquote {\bibinfo {title} {{T}racing {M}onopoles and {A}nti-monoploes in a
  {M}agnetic {H}edgehog {L}attice},}\ }\bibinfo {note}
  {{a}rXiv:1909.01316}\BibitemShut {NoStop}%
\bibitem [{\citenamefont {Binz}\ and\ \citenamefont
  {Vishwanath}(2008)}]{Binz2008}%
  \BibitemOpen
  \bibfield  {author} {\bibinfo {author} {\bibfnamefont {B.}~\bibnamefont
  {Binz}}\ and\ \bibinfo {author} {\bibfnamefont {A.}~\bibnamefont
  {Vishwanath}},\ }\bibfield  {title} {\enquote {\bibinfo {title} {Chirality
  induced anomalous-{H}all effect in helical spin crystals},}\ }\href@noop {}
  {\bibfield  {journal} {\bibinfo  {journal} {Physica B}\ }\textbf {\bibinfo
  {volume} {403}},\ \bibinfo {pages} {1336} (\bibinfo {year}
  {2008})}\BibitemShut {NoStop}%
\bibitem [{\citenamefont {Kakihana}\ \emph {et~al.}(2018)\citenamefont
  {Kakihana}, \citenamefont {Aoki}, \citenamefont {Nakamura}, \citenamefont
  {Honda}, \citenamefont {Nakashima}, \citenamefont {Amako}, \citenamefont
  {Nakamura}, \citenamefont {Sakakibara}, \citenamefont {Hedo}, \citenamefont
  {Nakama},\ and\ \citenamefont {^^c5^^8cnuki}}]{Kakihana2018}%
  \BibitemOpen
  \bibfield  {author} {\bibinfo {author} {\bibfnamefont {M.}~\bibnamefont
  {Kakihana}}, \bibinfo {author} {\bibfnamefont {D.}~\bibnamefont {Aoki}},
  \bibinfo {author} {\bibfnamefont {A.}~\bibnamefont {Nakamura}}, \bibinfo
  {author} {\bibfnamefont {F.}~\bibnamefont {Honda}}, \bibinfo {author}
  {\bibfnamefont {M.}~\bibnamefont {Nakashima}}, \bibinfo {author}
  {\bibfnamefont {Y.}~\bibnamefont {Amako}}, \bibinfo {author} {\bibfnamefont
  {S.}~\bibnamefont {Nakamura}}, \bibinfo {author} {\bibfnamefont
  {T.}~\bibnamefont {Sakakibara}}, \bibinfo {author} {\bibfnamefont
  {M.}~\bibnamefont {Hedo}}, \bibinfo {author} {\bibfnamefont {T.}~\bibnamefont
  {Nakama}}, \ and\ \bibinfo {author} {\bibfnamefont {Y.}~\bibnamefont
  {^^c5^^8cnuki}},\ }\bibfield  {title} {\enquote {\bibinfo {title} {Giant
  {H}all {R}esistivity and {M}agnetoresistance in {C}ubic {C}hiral
  {A}ntiferromagnet {E}u{P}t{S}i},}\ }\href@noop {} {\bibfield  {journal}
  {\bibinfo  {journal} {J. Phys. Soc. Japan}\ }\textbf {\bibinfo {volume}
  {87}},\ \bibinfo {pages} {023701} (\bibinfo {year} {2018})}\BibitemShut
  {NoStop}%
\bibitem [{\citenamefont {Kaneko}\ \emph {et~al.}(2019)\citenamefont {Kaneko},
  \citenamefont {Frontzek}, \citenamefont {Matsuda}, \citenamefont {Nakao},
  \citenamefont {Munakata}, \citenamefont {Ohhara}, \citenamefont {Kakihana},
  \citenamefont {Haga}, \citenamefont {Hedo}, \citenamefont {Nakama},\ and\
  \citenamefont {Onuki}}]{Kaneko2019}%
  \BibitemOpen
  \bibfield  {author} {\bibinfo {author} {\bibfnamefont {K.}~\bibnamefont
  {Kaneko}}, \bibinfo {author} {\bibfnamefont {M.~D.}\ \bibnamefont
  {Frontzek}}, \bibinfo {author} {\bibfnamefont {M.}~\bibnamefont {Matsuda}},
  \bibinfo {author} {\bibfnamefont {A.}~\bibnamefont {Nakao}}, \bibinfo
  {author} {\bibfnamefont {K.}~\bibnamefont {Munakata}}, \bibinfo {author}
  {\bibfnamefont {T.}~\bibnamefont {Ohhara}}, \bibinfo {author} {\bibfnamefont
  {M.}~\bibnamefont {Kakihana}}, \bibinfo {author} {\bibfnamefont
  {Y.}~\bibnamefont {Haga}}, \bibinfo {author} {\bibfnamefont {M.}~\bibnamefont
  {Hedo}}, \bibinfo {author} {\bibfnamefont {T.}~\bibnamefont {Nakama}}, \ and\
  \bibinfo {author} {\bibfnamefont {Y.}~\bibnamefont {Onuki}},\ }\bibfield
  {title} {\enquote {\bibinfo {title} {Unique {H}elical {M}agnetic {O}rder and
  {F}ield-{I}nduced {P}hase in {T}rillium {L}attice {A}ntiferromagnet
  {E}u{P}t{S}i},}\ }\href@noop {} {\bibfield  {journal} {\bibinfo  {journal}
  {J. Phys. Soc. Japan}\ }\textbf {\bibinfo {volume} {88}},\ \bibinfo {pages}
  {013702} (\bibinfo {year} {2019})}\BibitemShut {NoStop}%
\bibitem [{\citenamefont {Takeuchi}\ \emph {et~al.}(2019)\citenamefont
  {Takeuchi}, \citenamefont {Kakihana}, \citenamefont {Hedo}, \citenamefont
  {Nakama},\ and\ \citenamefont {Onuki}}]{Takeuchi2019}%
  \BibitemOpen
  \bibfield  {author} {\bibinfo {author} {\bibfnamefont {T.}~\bibnamefont
  {Takeuchi}}, \bibinfo {author} {\bibfnamefont {M.}~\bibnamefont {Kakihana}},
  \bibinfo {author} {\bibfnamefont {M.}~\bibnamefont {Hedo}}, \bibinfo {author}
  {\bibfnamefont {T.}~\bibnamefont {Nakama}}, \ and\ \bibinfo {author}
  {\bibfnamefont {Y.}~\bibnamefont {Onuki}},\ }\bibfield  {title} {\enquote
  {\bibinfo {title} {Magnetic {F}ield versus {T}emperature {P}hase {D}iagram
  for ${H}\parallel$ [001] in the {T}rillium {L}attice {A}ntiferromagnet
  {E}u{P}t{S}i},}\ }\href@noop {} {\bibfield  {journal} {\bibinfo  {journal}
  {J. Phys. Soc. Japan}\ }\textbf {\bibinfo {volume} {88}},\ \bibinfo {pages}
  {053703} (\bibinfo {year} {2019})}\BibitemShut {NoStop}%
\bibitem [{\citenamefont {Kurumaji}\ \emph {et~al.}(2019)\citenamefont
  {Kurumaji}, \citenamefont {Nakajima}, \citenamefont {Hirschberger},
  \citenamefont {Kikkawa}, \citenamefont {Yamasaki}, \citenamefont {Sagayama},
  \citenamefont {Nakao}, \citenamefont {Taguchi}, \citenamefont {Arima},\ and\
  \citenamefont {Tokura}}]{Kurumaji2019}%
  \BibitemOpen
  \bibfield  {author} {\bibinfo {author} {\bibfnamefont {Takashi}\ \bibnamefont
  {Kurumaji}}, \bibinfo {author} {\bibfnamefont {Taro}\ \bibnamefont
  {Nakajima}}, \bibinfo {author} {\bibfnamefont {Max}\ \bibnamefont
  {Hirschberger}}, \bibinfo {author} {\bibfnamefont {Akiko}\ \bibnamefont
  {Kikkawa}}, \bibinfo {author} {\bibfnamefont {Yuichi}\ \bibnamefont
  {Yamasaki}}, \bibinfo {author} {\bibfnamefont {Hajime}\ \bibnamefont
  {Sagayama}}, \bibinfo {author} {\bibfnamefont {Hironori}\ \bibnamefont
  {Nakao}}, \bibinfo {author} {\bibfnamefont {Yasujiro}\ \bibnamefont
  {Taguchi}}, \bibinfo {author} {\bibfnamefont {Taka-hisa}\ \bibnamefont
  {Arima}}, \ and\ \bibinfo {author} {\bibfnamefont {Yoshinori}\ \bibnamefont
  {Tokura}},\ }\bibfield  {title} {\enquote {\bibinfo {title} {Skyrmion lattice
  with a giant topological {H}all effect in a frustrated triangular-lattice
  magnet},}\ }\href {\doibase 10.1126/science.aau0968} {\bibfield  {journal}
  {\bibinfo  {journal} {Science}\ }\textbf {\bibinfo {volume} {365}},\ \bibinfo
  {pages} {914--918} (\bibinfo {year} {2019})}\BibitemShut {NoStop}%
\end{thebibliography}%

\end{document}